\documentclass[pre,superscriptaddress,twocolumn,nofootinbib,showpacs,floatfix]{revtex4}
\usepackage{amsmath,amssymb,graphicx,tikz,simplewick,verbatim,color,hyperref,cancel}
\usepackage{calc}
\usepackage{array}
\usepackage{bm}
\usepackage[percent]{overpic}
\newcommand{\aone}{magenta}
\newcommand{\atwo}{blue}
\newcommand{\athree}{red}
\newcommand{\maone}{green}
\newcommand{\matwo}{orange}
\newcommand{\mathree}{cyan}

\usepackage{shorttitles}

\newcommand{\dbedit}[1]{#1}
\begin{document}

\title{Plastic deformation of tubular crystals by dislocation glide}

\author{Daniel A.\ Beller}  
\affiliation{John A.\ Paulson School of Engineering and Applied Sciences, Harvard University, Cambridge, MA 02138, USA}
\author{David R.\ Nelson}  
\affiliation{Department of Physics, Harvard University, Cambridge, Massachusetts 02138, USA}

\date{\today}

\begin{abstract}
Tubular crystals, two-dimensional lattices wrapped into cylindrical topologies, arise in many contexts, including botany and biofilaments, and in physical systems such as carbon nanotubes. The geometrical principles of botanical phyllotaxis, describing the spiral packings on cylinders commonly found in nature, have found application in all these systems. Several recent studies have examined defects in tubular crystals associated with crystalline packings that must accommodate a fixed tube radius. Here, we study the mechanics of tubular crystals with variable tube radius, with dislocations interposed between regions of different phyllotactic packings. Unbinding and separation of dislocation pairs with equal and opposite Burgers vectors allow the growth of one phyllotactic domain at the expense of another. In particular, glide separation of dislocations offers a low-energy mode for plastic deformations of solid tubes in response to external stresses, reconfiguring the lattice step by step. Through theory and simulation, we examine how the tube's radius and helicity affects, and is in turn altered by, the mechanics of dislocation glide. We also discuss how a sufficiently strong bending rigidity can alter or arrest the deformations of tubes with small radii.\end{abstract}

\maketitle


\section{Introduction}

\begin{figure}[htbp]
\includegraphics[width=\linewidth]{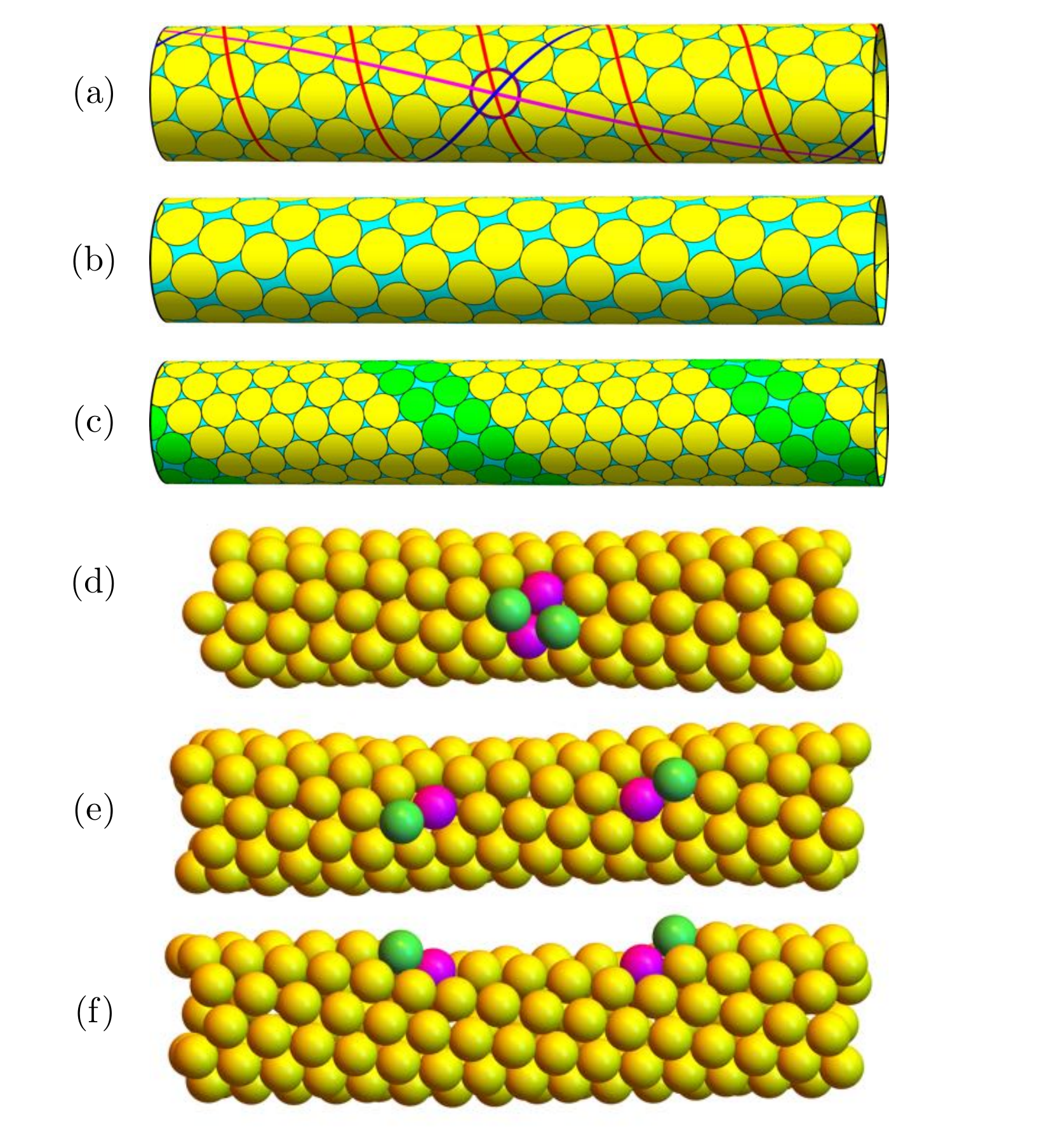}
    \caption{\dbedit{(a) A triangular lattice of disks in a cylindrical surface. The three parastichies intersecting at the disk outlined in purple are shown as the \aone, \atwo, and \athree\ helices. (b) A rhombic packing of disks in a cylindrical surface. (c) A triangular packing of disks in a cylindrical surface with a spiral  line-slip defect or ``stacking fault''. The disks in contact with the line-slip defect have reduced coordination number (5 neighbors instead of 6) and are colored green. (d) A tubular crystal consisting of a triangular packing of spheres (yellow) interrupted by the nucleation of a dislocation pair, producing two five-coordinated (green) and two seven-coordinated (magenta) spheres. (e) The dislocations nucleated in (d) are now spatially separated. (f) A different view of the tubular crystal in (e), highlighting that the region between the dislocations is narrower than the region outside the dislocations. }  \label{disksfig}  }
\end{figure}

In botany, the arrangements of leaves around a stem, seeds on a pinecone, spines on a cactus, scales on a pineapple, etc.\  often follow beautifully regular spiraling patterns that have been a subject of interest for centuries, known as phyllotaxis (``leaf arrangement'') \cite{adler1997history,kuhlemeier2007phyllotaxis,pennybacker2015phyllotaxis}. The spirals of nearest-neighbor connections, known as parastichies, form families of parallel curves characterized by integer indices called parastichy numbers, indicating the number of distinct spirals in the family (see Fig.~\ref{disksfig}(a) and Fig.~\ref{introfig}(b)). An intriguing feature of botanical phyllotaxis has been the widespread appearance of parastichy numbers that are successive members of the Fibonacci sequence (or of  similar sequences called the double Fibonacci sequence and the Lucas sequence \cite{pennybacker2015phyllotaxis}). The resulting divergence angle $d$, the azimuthal angle between consecutive sites (ordered by radius or height),   is then approximately related to the golden mean $\varphi = \tfrac{1}{2}(1+\sqrt{5})$ by $d\approx 2\pi(1-\varphi^{-1})\approx 137.5^\circ$\cite{adler1997history,kuhlemeier2007phyllotaxis,pennybacker2015phyllotaxis}. 

In recent decades, phyllotaxis has been shown not to belong exclusively to botany; it can also arise in physical systems under the general scenario of isotropically repulsive particles self-organizing on a compressing cylinder or a growing disc. Levitov predicted that repulsive vortices in a Type II superconductor would naturally converge toward  Fibonacci parastichy numbers \cite{levitov1991phyllotaxis,levitov1991energetic}. Fibonacci spirals were in fact were soon observed in experiments on repulsive ferrofluid drops in a magnetic field \cite{douady1992phyllotaxis}. Recently, the dynamics of such phyllotactic growth have been studied in a ``magnetic cactus'' model of repulsive magnets on a cylinder \cite{PhysRevLett.102.186103,nisoli2010annealing}. 

\dbedit{More generally}, regular helical packings on cylinders that do not necessarily follow the Fibonacci or Lucas sequences are widespread in biology and physics. Such systems were called  ``tubular crystals'' by  Erickson \cite{erickson1973tubular}, who  suggested that the geometrical language of phyllotaxis, including the parastichy number labeling, is a natural description for such systems. He had in mind tubular assemblies in microbiology such as rod-like viruses or bacteriophage tails, bacterial flagella, and intracellular biofilaments such as actin and microtubules. Packings of spherical particles in cylindrical capillaries or on cylindrical surfaces also show more general phyllotactic arrangements \cite{PhysRevE.81.040401,mughal2011phyllotactic,wood2013self}. Covalently-bonded single-walled carbon nanotubes (SWCNTs) and related materials such as boron nitride nanotubes  provide further important examples: their hexagonal unit cells form helical lattice lines along the tube, with a traditional labeling by a pair of integer indices in correspondence with the parastichy numbers of phyllotaxis \cite{harris2009carbon}. 

In this paper, we study the mechanics of plastic deformation of tubular crystals via the nucleation and glide separation of pairs of dislocation defects in the tubular lattice. The motion of these dislocations causes a \textit{parastichy transition}, i.e., a change in  the parastichy numbers of the tubular crystal \cite{harris1980tubular}, thus providing a low-energy mode for the release of externally imposed strain. For maximum simplicity, we focus on  triangular lattices, which are described by an isotropic elastic tensor, and treat the tubes as thin-sheet materials. In simulations, we model these materials as networks of harmonic springs with a bending rigidity. In the continuum limit, the elastic interactions of dislocations on  perfectly cylindrical surfaces of fixed radius have recently been studied \cite{amir2013theory}. Here we focus on  the changes in tube radius and helicity that accompany the parastichy transitions for a \dbedit{finite-sized elastic network} subject to plastic deformations.

When disks are packed on a cylindrical surface of fixed radius $R$, incommensurability of a perfect triangular lattice with the azimuthal periodicity $2\pi R$ can cause the ground state either to distort into a strained rhombic lattice, \dbedit{such as the one depicted in Fig.~\ref{disksfig}(b)}\cite{van1907mathematische,harris1980tubular}, or else to develop a one-dimensional, helical ``line-slip'' defect interrupting an otherwise triangular packing \dbedit{like the one in Fig.~\ref{disksfig}(c)} \cite{mughal2014theory}. The same is true for packings of spheres that are constrained to lie in contact with a solid cylinder, as the sphere centers all sit at a fixed radius $R$ from the center line \cite{mughal2012dense,mughal2013screw,wood2013self}. The relative stability of uniform rhombic packings versus line-slip packings has been shown recently to depend on the softness of the interparticle potential \cite{wood2013self}.

Our focus here is on a different kind of system: Although we retain the tube topology, we remove the constraint of a perfectly cylindrical substrate or wall of fixed radius, instead allowing the tubular crystal's shape to vary as a function of space and time. The tube shape is determined by the energy-minimizing positions of sites with a given bond network, possibly including defects.  By construction, in this system there are no line-slip defects  or extended rhombic packings. Instead, the triangular lattice topology is interrupted only by isolated dislocations nucleating in pairs, as depicted in Fig.~\ref{disksfig}(d), and moving by successive bond flips (see Fig.~\ref{bondflipfig}) to produce intermediate states like the one shown in Fig.~\ref{disksfig}(e,f). The total number of bonds in the lattice is conserved in this process. The tube radius and the orientation of the lattice on the tube both adjust in response to  the passage of a dislocation through the system, due to well-established geometrical rules that we describe in Section \ref{phyllsec}. For example, the tubular crystal in Fig.~\ref{disksfig}(e,f) is narrower in the region between the dislocations than in the region outside. Further motion of the dislocations away from each other expands the new, narrower tessellation at the expense of the original, wider one, resulting in a plastically deformed tubular crystal.

These considerations describe SWCNTs, where the honeycomb lattice of hexagons can be interrupted by dislocations comprised of pentagon-heptagon pairs. Such dislocations can arise and  move through the lattice via successive carbon-carbon bond rotations, called Stone-Wales rotations, altering the tube's radius and helicity in the process \cite{yakobson1998mechanical}. Detailed quantum mechanical simulations have been employed to study  plastic deformation of these objects by dislocation motion  \cite{PhysRevLett.81.4656,nardelli1998mechanism,yakobson2001mechanical,zhang2009dislocation,bettinger2002mechanically}, and there is experimental evidence confirming the role of defect motion in the plasticity of strained nanotubes \cite{huang2006kink,bozovic2003plastic,suenaga2007imaging}.  One reason for the importance of such dislocation-mediated deformations is that the resulting change in phyllotactic indices alters the SWCNT's electronic properties \cite{yakobson1998mechanical,yakobson2001mechanical}. In addition, the presence of dislocations within a SWCNT provides a source of disorder that  decreases the conductivity \cite{nardelli1999mechanical,bozovic2003plastic}. 

Microtubules offer another example of helical tubular crystals, in this case composed of tubulin proteins, where the tube radius is determined by the crystalline network. Microtubules are biofilaments responsible for the mechanics of the eukaryotic cytoskeleton, and are especially important in cell division \cite{nogales2000structural}. Here, dimers of the protein tubulin assemble into protofilaments parallel or nearly parallel to the tube axis, with an axial shift between adjacent protofilaments giving rise to the shallower helical parastichy \cite{hunyadi2007microtubule}. Microtubules  have been observed to change protofilament number along their length, strongly suggesting the presence of dislocations \cite{chretien1992lattice}. 
 
Dislocation-mediated plastic deformation has also been suggested as a mechanism for the growth of rod-shaped bacteria such as \textit{E.\ coli}, via circumferential climb motion of dislocations  in the peptidoglycan network of the bacterial cell wall, with new material added to the wall at each climb step \cite{nelson2012biophysical,amir2012dislocation,nelson2013defects}. Plastic bending deformations of growing bacteria can also be understood in the context of this model  \cite{amir2014bending}. 

The rest of this paper is organized as follows. In Section \ref{phyllsec} we describe the phyllotactic geometry of tubular crystals and the parastichy transformations caused by the unbinding of dislocations. Section \ref{sec3}  gives the predictions of continuum elasticity for the mechanics of plastic deformation in tubular crystals, first without a bending energy and then with a bending energy introduced as an important perturbative correction. Section \ref{secnum} presents results of numerical simulations modeling tubular crystals as networks of harmonic springs with a bending energy, probing the critical axial tension required to unbind and separate dislocation pairs over a wide range of tube geometries. Finally, in Section \ref{necksec} we examine the slight narrowing  of the tube ``neck'' around an isolated dislocation, as the radius of the tube changes abruptly but also exhibits interesting oscillatory behavior.

\section{Phyllotactic description of tubular crystals \label{phyllsec}}

\begin{figure}[htbp]
\includegraphics[width=\linewidth]{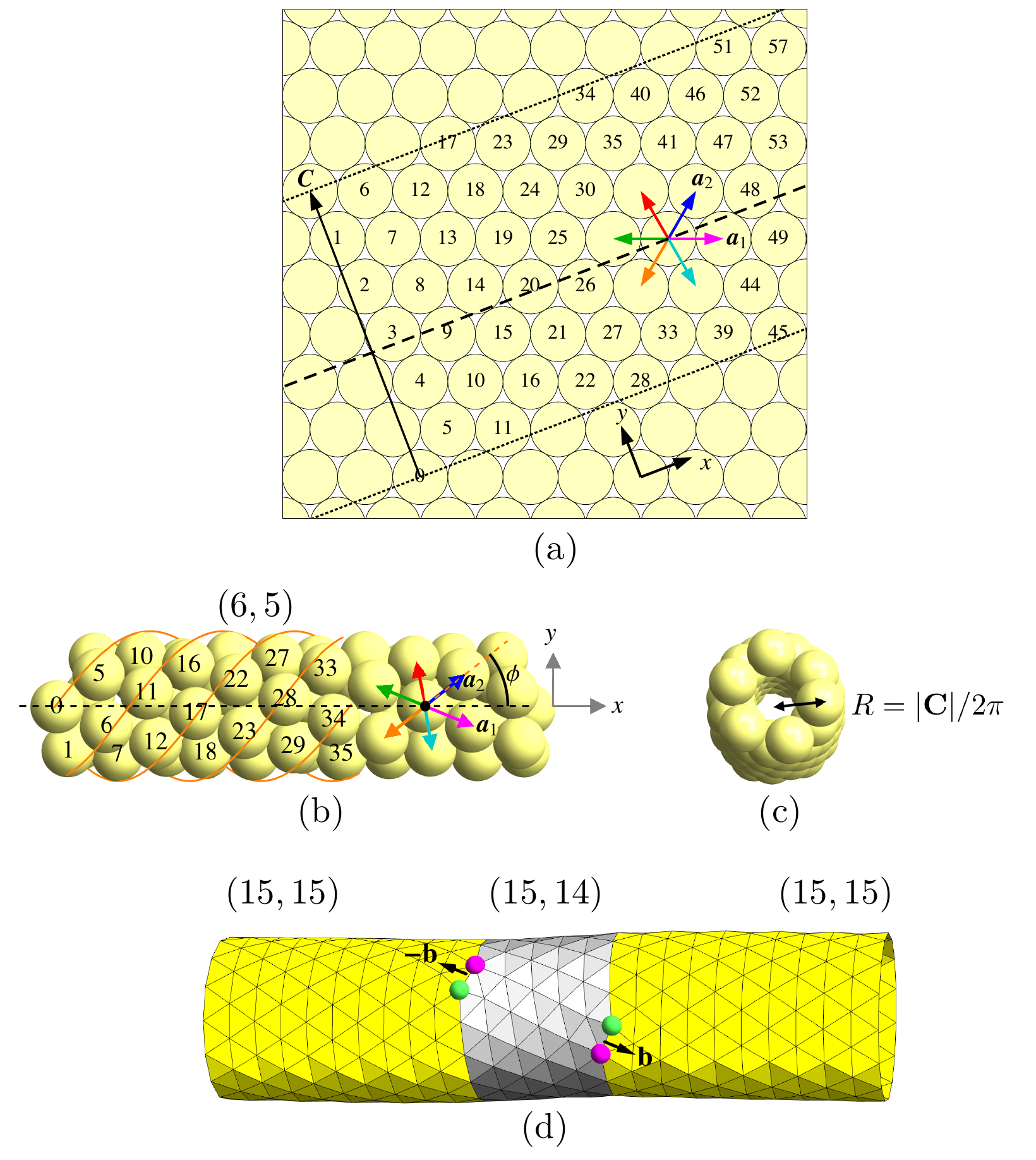}
\caption{ Schematic illustration of the phyllotactic arrangement of sites in tubular triangular crystals.  (a) 2D triangular packing of discs, with a circumference vector $\mathbf C$ selected to roll the planar lattice into a cylinder with phyllotactic indices $(m,n)=(6,5)$ by identifying the two dotted lines, giving a cylinder axis parallel to the dashed line. The six primitive lattice directions are shown as the six colored arrows. Number labels are the axial index of the lattice sites in order of increasing $x$-coordinate (coordinate axes shown in gray). (b,c) The tubular crystal with $(m,n)=(6,5)$, a triangular packing of spheres approximating the cylindrical packing of discs obtained from (a), in both side (b) and top (c) views. The geometrical parameters $\phi$, describing the lattice orientation, and $R$, the tube radius, are shown in (b) and (c) respectively. Orange helices in (b) follow portions of the 5 parastichies along $\bf a_2$. (d) Plastic deformation of a tubular crystal from an initial $(m,n)=(15,15)$ configuration (yellow) to $(m',n')=(15,14)$ (gray) by glide of a dislocation pair with $\mathbf b = \mathbf a_1$; the figure shows a snapshot of the deformation process as the dislocations glide apart. Positive and negative disclinations are marked by green and magenta spheres, respectively.\label{introfig}    }
\end{figure}

In the study of phyllotaxis, the helical arrangements of leaves, petals, etc.\ are described in terms of the helices along directions of nearest-neighbor contacts, called \textit{parastichies} \cite{rothen1989phyllotaxis}. These contacts define lattice directions of an associated 2D lattice that has been rolled into a cylinder. Each parastichy is a member of a family of stacked parastichy helices, related by translations along the cylinder axis. \dbedit{Fig.~\ref{disksfig}(a) shows  three parastichies threading through a particular lattice site in a triangular lattice, while the orange helices in Fig.~\ref{introfig}(b) comprise the parastichy family consisting of all parastichies along the direction  $\pm\mathbf a_2$ as labeled in the figure.}  Each parastichy family collectively accounts for all lattice sites. The number of distinct members of a particular parastichy family gives a \textit{parastichy number}. If all the lattice sites are ordered by their height along the cylinder's axis and labeled with an \textit{axial index} $i$, then two sites neighboring along a parastichy of parastichy number $q$ will differ in their axial indices by $\Delta i = q$ \cite{rothen1989phyllotaxis}.

For a triangular lattice, there are three families of parastichies, and the tubular crystal topology is described by a triple of integer parastichy numbers $(|n-m|,n,m)$. If all three (positive) parastichy numbers have a common factor $k>1$, then the lattice has a $k$-fold rotational symmetry, where $k$ is known as the \textit{jugacy}.  Since the first of the three parastichy numbers is the difference of the other two, a \dbedit{triangular lattice tessellation of a tubular crystal} can be uniquely labeled by an ordered pair of parastichy numbers. We choose to use the representation $(m,n)$ where $m$ is the parastichy number of the steepest right-handed helix, and $n$ that of the steepest left-handed helix. The indices are then restricted to the range $\frac{1}{2} n \leq m < 2n$. 

The geometry of a pristine tubular packing of spheres or the hexagons of a carbon nanotube is well-approximated by a triangular packing of discs in the surface of a cylinder \cite{harris1980tubular,erickson1973tubular}; and the latter situation is much easier to describe with precision. Consider a triangular packing of discs in the plane, with two primitive lattice vectors $\mathbf a_1$ and $\mathbf a_2$, where $\mathbf a_2$ is oriented $60^\circ$  counterclockwise from $\mathbf a_1$ and both vectors have length equal to the lattice spacing $a$. As shown in Fig.~\ref{introfig}, we can roll up the packing  into a tube by choosing a lattice vector $\mathbf C = c_1 \mathbf a_1 + c_2 \mathbf a_2$, with $c_1,c_2\in\mathbb{Z}$, to serve as a circumference vector (also referred to as the characteristic vector), so that lattice sites separated by $\mathbf C$ in the plane are mapped to the same site on the cylinder. Since every step upward along the $n$-parastichy raises the axial index by $\Delta i=n$, and likewise every step downward along the $m$-parastichy lowers the axial index by $\Delta i = -m$, the circumference vector can be expressed as $\mathbf C = - n \mathbf a_1 + m \mathbf a_2$. The resultant path has a net change of $\Delta i = -nm + mn = 0$ in the axial index, as must be true for a path beginning and ending at the same point on the cylinder. The radius of the tube (Fig.~\ref{introfig}(c)) is therefore, to an excellent approximation in all but the smallest-diameter tubes,
\begin{align}
R &\approx \frac{1}{2\pi} |\mathbf C| = \frac{a}{2\pi} \sqrt{m^2+n^2-mn} \label{Rdef}.
\end{align}
The orientation of the lattice on the tube can be described by the angle $\phi$ that the $n$-parastichy, the steepest left-handed helix of nearest neighbors, makes with the cylinder axis:
\begin{align}
\tan \phi &\approx \frac{2}{\sqrt{3}} \left(\frac{m}{n}-\frac{1}{2}\right). \label{phidef}
\end{align}
Using the $n$-parastichy to define $\phi$ is of course arbitrary; in general the lattice directions make angles $\phi + s \pi/3$, $s\in \mathbb{Z}$, with the cylinder axis. For a pristine (defect-free) tubular crystal in mechanical equilibrium, the pair of ``geometrical descriptors'' $(R,\phi)$ is thus in one-to-one correspondence with the pair of parastichy numbers $(m,n)$ that identifies the tesselation of the tubular crystal. Our choice $\frac{1}{2} n \leq m < 2n $ restricts $\phi$ to  the interval $0 \leq \phi < 60^\circ$. Figs.~\ref{introfig}(a,b) introduce a color-code for the six primitive lattice directions that we will use throughout this paper. \dbedit{Equations \ref{Rdef} and \ref{phidef} are exact for a packing of disks in a cylindrical surface, and nearly exact for all but the few smallest tubular crystals; more complicated, implicit equations for the geometrical parameters of a tubular crystal may be solved numerically to give $R$ and $\phi$ exactly \cite{erickson1973tubular,harris1980tubular}.} 

Two $\phi$-values correspond to achiral packings. For $\phi=0$, we have $m=\frac{1}{2} n$ and one of the parastichies is a straight line along the cylinder axis. The ``armchair'' single-walled carbon nanotube geometry has its hexagons centered on such a lattice \cite{harris2009carbon}. The other achiral configuration has $\phi=30^\circ$, $m=n$ and one of the parastichies is a circle around the cylinder's circumference; this corresponds to the hexagon centers in the ``zigzag'' nanotube geometry.

If a tube is to undergo plastic deformation changing $R$ and $\phi$, how can we accomplish the necessary \textit{parastichy transition} altering $(m,n)$? One option, termed ``continuous contraction'' by Harris and Erickson \cite{harris1980tubular}, is to break contacts along one of the parastichies uniformly throughout the lattice, thus replacing the triangular packing with a rhombic one (see Fig.~\ref{disksfig}(b)), and to distort the rhombic packing until a new set of contacts is made, reestablishing the triangular lattice with a new pair of parastichy numbers \cite{van1907mathematische}. \dbedit{Such continuous contractions are the basis for parastichy transitions in botany and are essential to understanding the widespread appearance of numbers from the Fibonacci and similar sequences as parastichy numbers \cite{kuhlemeier2007phyllotaxis,pennybacker2015phyllotaxis}}. Levitov studied continuous transitions for two-dimensional lattices of repulsive particles under compression, finding a physical basis for the prevalence of Fibonacci parastichy numbers in this scenario \cite{levitov1991phyllotaxis,levitov1991energetic}. However, \dbedit{in a physical context,} such transitions typically involve large energy barriers, proportional to the length of the parastichy with broken bonds.

Here we instead examine a much more localized (and lower energy) type of parastichy transition: the motion of a dislocation pair through the triangular lattice. Each dislocation can be viewed as a pair of point disclinations, a positive disclination at a five-coordinated site and a negative disclination at a seven-coordinated site (see, e.g., \cite{Nelson}). The dislocation is characterized by a Burgers vector $\mathbf b = - \oint (\partial \mathbf u /\partial l) dl $ where $\mathbf u$ is the vector field of displacements from the pristine lattice, and the integration is along a counterclockwise Burgers circuit enclosing the dislocation \cite{landau1975elasticity}. From a pristine triangular lattice, a pair of dislocations with Burgers vectors $\mathbf b$, $-\mathbf b$ can be created by a bond flip, transforming four six-coordinated sites to a pair of 5-7 disclination pairs (see Fig.~\ref{bondflipfig}). Here $\mathbf b$ is a two-dimensional vector, $\mathbf b = b_1 \mathbf a_1+b_2 \mathbf a_2$, with $b_1,b_2\in\mathbb{Z}$.

The dislocations can then mediate plastic deformations by moving through the lattice by either glide (along the directions $\pm \mathbf b$) or climb (perpendicular to $\mathbf b$) via successive local rearrangements. As the pair of dislocations moves apart, a region of tubular crystal with a different circumference vector $\mathbf C' = \mathbf C + \mathbf b$, grows between the two dislocations, at the expense of the original tubular topology $\mathbf C$ \cite{harris1980tubular}. Following the ``Frank criterion'' \cite{hirth1982theory}, we will consider only dislocations with Burgers vectors equal to the primitive lattice vectors, $\mathbf b = \pm \mathbf a_1$, $\pm \mathbf a_2$, or $\pm \mathbf a_3$ where $\mathbf a_3 = \mathbf a_2-\mathbf a_1$. Therefore, each parastichy transition  changes each of $m,n$ by either $\pm 1$ or $0$, as given in the following table:

\begin{equation} \begin{array}{| c | c c c c c c |} 
\hline
\mathbf b & \mathbf a_1 & \mathbf a_2 & \mathbf a_3  &- \mathbf a_1 & -\mathbf a_2 & -\mathbf a_3 \\
\hline
\Delta m & 0 & +1 & +1 & 0 & -1 & -1 \\
\Delta n & -1 & 0 & +1 & +1 & 0 & -1 \\
\hline
\end{array}
\label{bmntable}
\end{equation}
\dbedit{For comparison, the high-energy barrier, continuous contraction process described above results in one of the (usually more drastic) parastichy transitions $(\Delta m, \Delta n ) \in \left\{ (-n,0), (0,m), (m-n,m-n)\right\}$ if $m> n$, or $(\Delta m, \Delta n ) \in \left\{(0,-m), (n,0), (n-m,n-m)\right\}$ if $n< m$  \cite{harris1980tubular}. (If $m=n$ then the only available continuous contraction doubles $n$.)}

Fig.~\ref{introfig}(d) illustrates the plastic deformation of a tubular crystal with initial parastichy numbers $(m,n)=(15,15)$ (yellow region) into a new tessellation $(m',n')=(15,14)$ (gray region) by glide motion of a dislocation pair whose right-moving dislocation has $\mathbf b = \mathbf a_1$. The central gray region is slightly narrower than the yellow region \dbedit{(as in Fig.~\ref{disksfig}(e,f))}, and the purely circumferential parastichies in the yellow region become a single shallow helix in the gray region.  The lattice sites of the tubular crystal are here the nodes of a discretized surface, as simulated using the method described below in Section \ref{secnum}.

We will restrict our attention to dislocation glide and not climb. Glide is more relevant to colloidal cylindrical crystals, provided the density of vacancies and interstitials is low, and exit and entry of colloidal particles to and from the surrounding medium are rare events.  Climb dynamics is more natural  in the context of elongating bacteria, where the extra material needed for climb comes from metabolic processes inside the cell \cite{amir2012dislocation}. Because a dislocation can glide either parallel or antiparallel to its Burgers vector, we have a total of six distinct Burgers vector pairs $\pm \mathbf b$ to consider for each $(m,n)$.

\section{Mechanics of plastic deformation: Analytic predictions \label{sec3}}

Our goal is to understand the mechanics of plastic deformation in tubular crystals, as mediated by dislocation glide. We start with the usual  elastic free energy for isotropic two-dimensional, planar crystals,
\begin{align}
F_s &= \tfrac{1}{2} \int dA \left(2 \mu u_{ij} u_{ij} + \lambda u_{kk}^2\right), \label{Lame}
\end{align}
where the six-fold symmetry of a triangular lattice is sufficient to ensure isotropy  \cite{landau1975elasticity}. Here $\mu$ and $\lambda$ are the Lam\'e coefficients,  and $u_{ij}(\mathbf{x}) =  \frac{1}{2} \left[\partial_i u_j(\mathbf{x}) + \partial_j u_i(\mathbf{x})\right]$ is the strain tensor in terms of the displacement vector field $\mathbf u(\mathbf{x})$. The two-dimensional Young's modulus is $Y=4\mu (\mu+\lambda)/\left(2\mu+\lambda\right)$ \cite{seung1988defects}. 

 In a stress tensor field $\sigma_{ij}(\bf x)$, a dislocation with Burgers vector $\mathbf b$ experiences a force
\begin{align} 
f_i &= b_k \sigma_{jk} \epsilon_{ijz},
\end{align}
known as the Peach-Kohler force \cite{landau1975elasticity,hirth1982theory}, where $\epsilon_{ijz}$ is the Levi-Civita tensor. On an infinite plane (as opposed to a cylinder), the interaction energy of two  dislocations with opposite Burgers vectors $\mathbf b $, $-\mathbf b$ gliding apart with separation $\mathbf r=r \mathbf{b}/|\mathbf{b}|$  is  \cite{nelson1979dislocation,hirth1982theory}
\begin{align} F_s(\mathbf r) &= A  |\mathbf{b}|^2 \ln(r/a)  - b_k \sigma^\text{ext}_{jk}\epsilon_{ijz} r_i + 2 E_c  
\end{align}
where $A \equiv Y/(4\pi)$, $\sigma^\text{ext}_{jk} $ is a constant external stress tensor field, $a$ is the lattice constant, and  $E_c$ is the core energy of a dislocation. Let $\theta$ be the angle between $\mathbf b$ and a preferred direction $\bm{\hat x}$ (this will ultimately become the cylinder axis; see Fig.~\ref{introfig}(a,b)). Under the Frank criterion $|\mathbf b| = a$,  the elastic energy reduces to 
\begin{align} F_s(r) & = A a^2 \ln(r/a)  + \frac{1}{2}  \left(\sigma^\text{ext}_{xx}-\sigma^\text{ext}_{yy}\right) s a r \sin(2\theta) \nonumber  \\
&\quad  - \sigma^\text{ext}_{xy} sar \cos(2\theta) + \text{const.}\; , \label{Fsplane}
\end{align}
where $s\equiv \text{sign}\left[\mathbf b \cdot \bm{\hat x}\right]$, which is $+1$ (or $-1$) if the dislocations increase their separation by gliding parallel (or antiparallel) to their Burgers vectors. We take the Burgers vector $\mathbf b$ to belong to the dislocation gliding with projection along the $+\bm{\hat x}$ direction. The corresponding  force along the glide direction is then
\begin{align} f_g(r) &= - A \frac{a^2}{r}  - \frac{1}{2} \left(\sigma^\text{ext}_{xx}-\sigma^\text{ext}_{yy}\right)s a \sin(2\theta) \nonumber \\
&\quad + \sigma^\text{ext}_{xy} s a \cos(2\theta).
\end{align}
At small separation $r$, the force is dominated by the attractive stress field of the dislocation pair. At large $r$, however, the dislocations can be driven apart, depending on the signs of the components of $\sigma^{\text{ext}}$ and the value of $\theta$. In this case, the force vanishes at a maximum in the one-dimensional energy landscape \cite{bruinsma1982motion,amir2013theory}, at separation
\begin{align} r^* &=  \frac{s Aa }{\sigma_{xy}^\text{ext} \cos(2\theta)-\frac{1}{2}\left(\sigma^\text{ext}_{xx}-\sigma^\text{ext}_{yy}\right) \sin(2\theta) }. \label{rstar} \end{align}
If we set $r^*$ equal to the lattice constant $a$, we can find the critical stress values necessary to pull apart a dislocation pair nucleated from the pristine lattice by a single bond flip (see Fig.~\ref{disksfig}(d) or Fig.~\ref{bondflipfig}(b)). These critical stresses, which we denote with a dagger symbol, are given by 
\begin{align}
\left(\sigma_{xx}-\sigma_{yy}\right)^\dagger &= -2sA/\sin(2\theta) , \label{sxxyydagger}\\
\sigma_{xy}^\dagger &= sA/\cos(2\theta)  \label{sxydagger}.
\end{align}
\dbedit{While continuum elasticity theory is expected to break down at the scale of the lattice constant $a$, we will find in Sec.~\ref{secnum} that continuum theory predictions obtained by setting $r^*=a$ nevertheless yield quantitative insights into numerically modeled  tubular crystals.   }

On a cylinder, the energetics of a dislocation pair at finite glide separation is different than on the plane, as each dislocation feels the stress field of all the periodic images of the other dislocation, so the dislocation pair interacts like a pair of grain boundaries in the infinite plane. The dislocation pair energy on a cylinder  was calculated in Ref.~\cite{amir2013theory}, and for a  separation $x$ along the cylinder axis and $y$ along the azimuthal direction can be written (up to a constant for fixed $\theta$) as
\begin{align}
\frac{F_s(x,y)}{A a^2/2}  &= \ln \left[ \cosh \tilde x - \cos \tilde y\right] \nonumber \\
&\quad  + \tilde x \left(\frac{ \sinh \tilde x \cos (2\theta) + \sin \tilde y \sin (2\theta)}{\cos \tilde y - \cosh \tilde x}\right) \nonumber \\
&\quad  + \frac{2R}{a} \left[ \tilde \sigma_{xy}(\tilde y \sin \theta - \tilde x \cos \theta) - \tilde \sigma_{yy} \tilde x \sin \theta \right. \nonumber \\
&\hspace{1.5cm}\left. + \tilde \sigma_{xx} \tilde y \cos \theta \right] \label{apn_energy}
\end{align}
where we use dimensionless quantities $\tilde x\equiv x/R$, $\tilde y \equiv y/R$, and $\tilde \sigma_{ij} =  \sigma^{\text{ext}}_{ij}/A$. Thus, at separations large compared to the cylinder radius $R$, the dislocations attract with a linear potential (second term of Eq.~\ref{apn_energy}) that can compete with the linear term from the external stress. 

For dislocation separations small compared to the cylinder radius, $r \ll R$, the stretching energy  for dislocations on a cylinder reduces to that on a plane. Therefore, in examining the stress required to unbind and separate a dislocation pair from an initial separation of one lattice spacing, it is appropriate to start with the stretching energy of Eq.\ \ref{Fsplane}, giving the critical stresses in Eqs.\ \ref{sxxyydagger} and \ref{sxydagger}. Those critical stresses are of order $A=Y/(4\pi)$, whereas the stress magnitude required to ensure that dislocations continue to glide apart at large distance $x\rightarrow \infty$ is of smaller order $\sim (a/R) A$, as obtained from the form of Eq.~(\ref{apn_energy}) \cite{amir2013theory}. So, an imposed stress great enough to unbind a dislocation pair at separation $r=a$ is sufficient to ensure continued glide to infinite separation.

Our continuum approach to dislocation energetics neglects a small periodic Peierls potential arising from the discreteness of the underlying lattice, as well as the associated Peierls stress needed to overcome it \cite{hirth1982theory}. In this paper, we assume a small nonzero temperature sufficient to allow easy passage over these corrections to our continuum energy formulas. Even without a small temperature to help overcome the Peierls barrier, the size of the critical stresses $\sigma^\dagger \sim A$ alone may be sufficient, as the maximum Peierls stress is typically several orders of magnitude smaller than $\mu \sim A$  \cite{hirth1982theory}. 

The external stress typically  acts to pull apart three of the six possible dislocation pairs with elementary, opposite Burgers vectors, whereas the other three dislocation pairs are pushed together. The critical stress values of Eqs.~(\ref{sxxyydagger},\ref{sxydagger}) do not depend on $R$, as the energetics at these tight separations are the same as on the plane;  however, they do depend on the lattice orientation $\phi$ with respect to the cylinder axis through the parastichy angle $\theta = \phi + s \cdot 60^\circ$, $s\in\mathbb{Z}$. On the other hand, each dislocation unbinding event changes both $R$ and $\phi$. (Recall from Sec.~\ref{phyllsec} that we define $\phi$ as the angle made by the steepest left-handed parastichy with $\bm{\hat x}$, the cylinder axis; see Eq.~(\ref{phidef}).)

\begin{figure}[htbp]
\includegraphics[width=\linewidth]{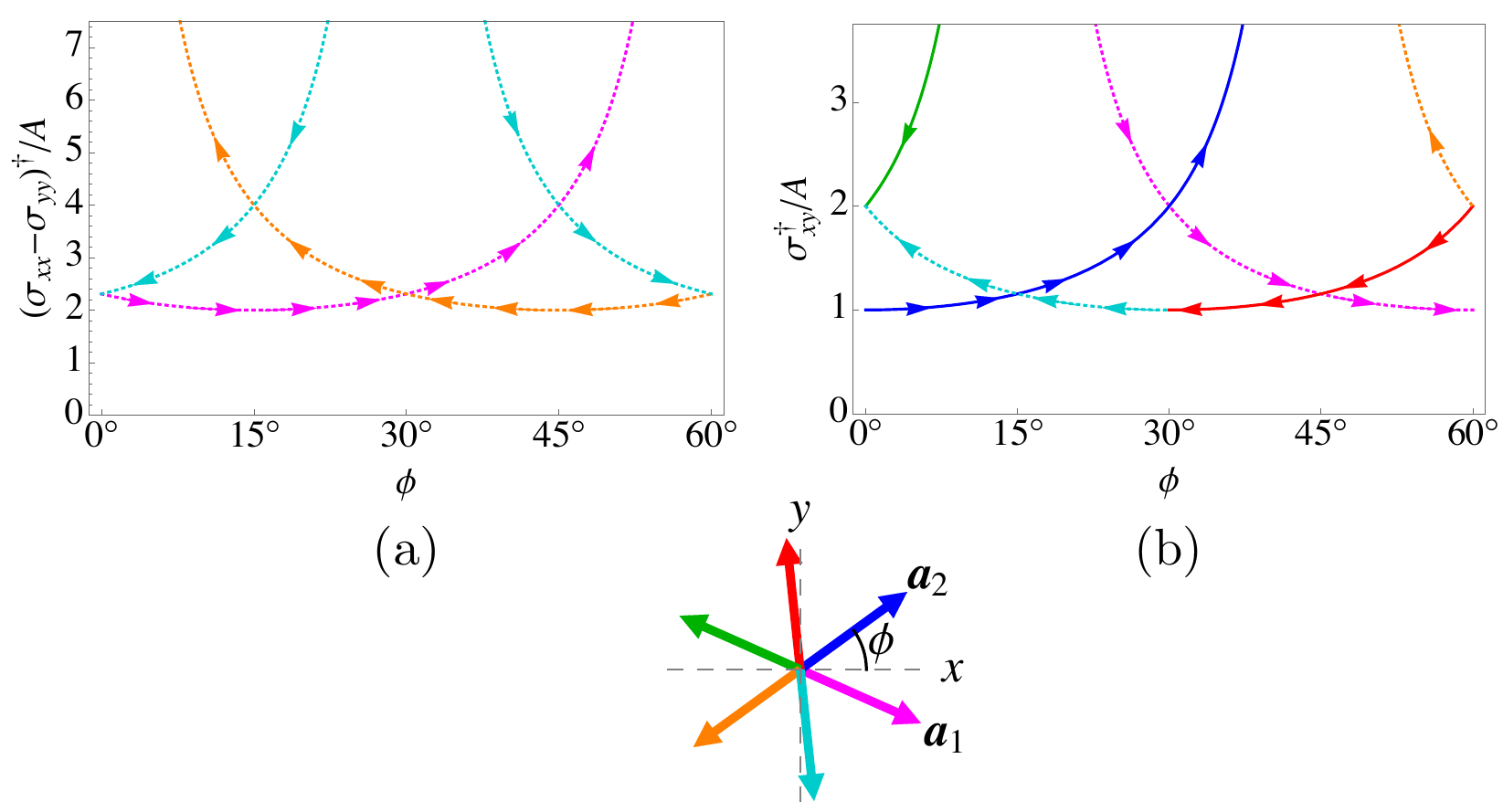}
\caption{Dimensionless critical external stress values $(\sigma_{xx}-\sigma_{yy})^\dagger/A$ (where $A=Y/4\pi$) and $\sigma_{xy}^\dagger/A$ required to unbind a dislocation pair with initial separation of one lattice spacing, for different lattice orientations $\phi$, as given in Eqs.\ \ref{sxxyydagger} and \ref{sxydagger}. The colored curves correspond to the right-moving dislocation (see Fig.~\ref{introfig}(d)) having the Burgers vector as depicted in the legend at bottom, where $\phi$ is measured relative to the cylinder axis $\bm{\hat x}$.  Arrowheads in (a) and (b)  indicate the direction of the change of helical angle $\phi$ with each plastic deformation event. Solid curves indicate that the dislocation unbinding increases $R$, whereas for dotted curves $R$ decreases. (a) Axial stress $\sigma_{xx}$ minus a pressure-like azimuthal stress $\sigma_{yy}$. (b) Torsional stress $\sigma_{xy}$. Note that all dislocation unbinding events in (a), with $\left(\sigma_{xx}-\sigma_{yy}\right)^\dagger>0$, decrease the tube radius $R$.} 
\label{sigmadaggertheory}
\end{figure}

The critical stresses of Eqs.~(\ref{sxxyydagger}) and (\ref{sxydagger}) for dislocation unbinding are plotted in Fig.~\ref{sigmadaggertheory} using the color scheme introduced in Fig.~\ref{introfig} for the Burgers vector $\mathbf b$. 
\dbedit{The arrows on the curves in Fig.~\ref{sigmadaggertheory}  record whether $\phi$ increases or decreases as a result of the associated  plastic deformation event.} If a tubular crystal with a given parastichy tilt $\phi$ is subjected to an external stress that slowly rises from zero until a dislocation unbinding event occurs, then we need only examine the Burgers vector pair with the lowest critical stress $\sigma^\dagger$ at a given $\phi$. In this scenario, for pure axial stress $\sigma^\dagger_{xx}>0$, we see in Fig.~\ref{sigmadaggertheory}(a)  a ``flow'' in $\phi$ away from $0$ and toward $30^\circ$, while the cylinder radius decreases. Because $\phi=30^\circ$ is an achiral geometry, the tube will thus evolve toward an (approximately) achiral state as its radius shrinks. This deformation pathway is in qualitative agreement with molecular dynamics simulations of carbon nanotubes showing that in the ductile regime of $\sim 10\%$ strain and high temperature, plastic deformations rotate the graphene orientation on the tube away from the armchair state ($\phi=0$) and toward the zigzag state ($\phi=30^\circ$) \cite{yakobson1998mechanical,PhysRevLett.81.4656}.

For an applied stress $\sigma_{xy}^\dagger$ that is purely torsional,  Fig.~\ref{sigmadaggertheory}(b) reveals a flow away from $\phi=45^\circ$ and toward $\phi=15^\circ$. If the sign of $\sigma_{xy}^{\text{ext}}$ is reversed, $\phi=15^\circ$ becomes unstable and $\phi=45^\circ$ becomes stable. Meanwhile, the radius changes nonmonotonically, depending on the value of $\phi$. Interestingly, molecular dynamics simulations of SWCNTs under torsion \cite{zhang2009dislocation} have found that $\phi=15^\circ$ is a critical angle at which there is a change in Burgers vector of the dislocation pair that becomes stable relative to the pristine lattice at lowest torsional strain, and at which the torsional strain per unit length $\gamma^*$ required to unbind this dislocation pair has a cusp qualitatively similar to that in Fig.~\ref{sigmadaggertheory}(b). Eq.~(\ref{sxydagger}) therefore appears to offer geometric intuition for those numerical findings. 

So far we have neglected the bending energy $F_b$ of the tubular crystal, which we now introduce:
\begin{align}
F_b &= \frac{1}{2} \kappa \int dA \left[H(\mathbf x)\right]^2 \label{Fb},
\end{align}
where $\kappa$ is the bending modulus and $H(\mathbf x)$ is the local mean curvature, equal to the sum of the two principal curvatures of the surface. There could in general be another term $\bar \kappa K(\mathbf{x})$ associated with the Gaussian curvature $K(\mathbf x)$, but we will restrict our attention to cylinders of infinite length (or with periodic boundary conditions) where the integrated Gaussian curvature vanishes. 

Comparing the Young's and bending moduli  gives rise to a length scale of local out-of-plane deformation $\sqrt{\kappa/Y}$ at which stretching and bending energies are of comparable magnitude. For cylinders of radius $R$, therefore, the relative importance of stretching and bending energies is characterized by a dimesionless ratio called the F\"oppl-von K\'arm\'an number (see, e.g., \cite{lidmar2003virus})
\begin{align}
\gamma &\equiv \frac{ Y R^2}{\kappa} .
\end{align}
For large $\gamma$, the tube will prefer to bend to minimize the elastic  energy; for small $\gamma$, local changes in the lattice constant leading to stretching or compression are the preferred mode of deformation. 

How does the bending energy affect the energetics of plastic deformation in a tube? For a pristine tubular crystal in the continuum approximation, with an exactly cylindrical shape of length $L$ and radius $R$, the bending energy approaches a simple limit,
\begin{align}
F_b &\rightarrow \pi \kappa L /R.
\end{align}
The bending energy is thus decreased by increasing the cylinder radius and decreasing its length; for glide deformations that keep the number of particles fixed, $F_b$  favors changing $m$ and $n$ to obtain larger $R(m,n)\approx (a/2\pi)\sqrt{m^2+n^2-mn}$. Now consider a tube for which the stretching energy $F_s$ has a local minimum with $L=L_0$ and $R=R_0$. If $\gamma=YR_0^2/\kappa$ is large but finite, we expect a slight increase in radius $R=R_0(1+u_{yy})$, $0< u_{yy}  \ll 1$, and a small decrease in length, $L=L_0(1+u_{xx})$, $u_{xx} < 0 $, $|u_{xx}|\ll 1$, where $u_{yy}$ is the azimuthal strain and $u_{xx}$ is the strain along the cylinder axis. Upon expanding in the small quantities $|u_{xx}|$, $u_{yy}$, and $\gamma^{-1}$, we find the total energy of the cylinder,\dbedit{
\begin{align}
F_{\text{tot}} &= F_b+F_s \nonumber \\
&= \frac{\pi \kappa L}{R} \nonumber \\
&\quad + \frac{1}{2}\cdot  \left(2\pi RL\right)\left[ 2\mu  \left(u_{xx}^2+u_{yy}^2\right) + \lambda  \left(u_{xx}+u_{yy}\right)^2 \right] \label{Ftotfull} \\
&\approx \pi  R_0 L_0 \left[ Y \gamma^{-1} \left(1+u_{xx}-u_{yy}\right) \right. \nonumber \\
&\quad \left. +  \left(2\mu+\lambda\right)\left(u_{xx}^2+u_{yy}^2\right) + 2 \lambda u_{xx} u_{yy} \right] . \label{Ftotapprox}
\end{align}
To first order in $\gamma^{-1}$, $F_{\text{tot}}$ is minimized by $u_{yy} = - u_{xx} = Y \gamma^{-1}/(4\mu) = \frac{1}{2} \left(1+\nu\right) \gamma^{-1}$, where $\nu = \lambda/(2\mu+\lambda)$ is the Poisson ratio \cite{seung1988defects}. } The bending energy therefore acts like a diagonal, traceless, radius-dependent contribution to the stress tensor,
\begin{align}
\sigma^b_{yy} &= - \sigma^b_{xx} = \frac{1}{2} Y\gamma^{-1} , \quad \quad \sigma^b_{xy} = 0 \label{sigmab} ,
\end{align}
as can be verified by replacing $F_b$ with $-  \int dA \sigma^b_{ij} u_{ij} = -\int dA \left(\sigma^b_{yy} u_{yy} +\sigma^b_{xx} u_{xx} \right) $ 
in $F_\text{tot}$. 

How does the bending energy affect the stresses required for plastic deformation by dislocation glide? As seen in Eq.~\ref{sxxyydagger}, the critical axial stress $\sigma_{xx}^\dagger$ and azimuthal stress $\sigma_{yy}^\dagger$ oppose one another, affecting glide motion only in the combination $(\sigma_{xx}-\sigma_{yy})^\dagger$. We have just found that the bending energy's effective stress contribution has $\sigma_{xx}^b = - \sigma_{yy}^b$. Therefore, the \textit{effective} critical stress contains a simple ``curvature offset'',
\begin{align}
\left(\sigma_{xx}-\sigma_{yy}\right)^{\dagger}_{ \text{eff}}  &= \left(\sigma_{xx}-\sigma_{yy}\right)^\dagger + \left(\sigma^b_{xx}-\sigma^b_{yy}\right) \nonumber \\
&=  \left(\sigma_{xx}-\sigma_{yy}\right)^\dagger - 4 \pi A \gamma^{-1} \label {curvatureoffset} .
\end{align}
The left-hand side, the effective critical stress to unbind a dislocation pair, must match the critical stress calculated in Eq.~(\ref{sxxyydagger}), which depends on $\phi$ but not $R$. Meanwhile, the curvature offset depends on $R$ but not $\phi$. The quantity $(\sigma_{xx}-\sigma_{yy})^\dagger$ on the right-hand side  of Eq.~(\ref{curvatureoffset}) is  the externally imposed stress actually required to unbind a dislocation pair in the presence of a bending rigidity. 

For the case of  imposed torsional stress, we can obtain a similar curvature-induced stress offset by replacing $\left(\sigma^\text{ext}_{xx}-\sigma^\text{ext}_{yy}\right)$ in Eq.~(\ref{rstar}) with $\left(\sigma_{xx}^b-\sigma_{yy}^b\right)= - 4\pi A\gamma^{-1}$ and solving for $\sigma_{xy}^{\text{ext}}$ to obtain $(\sigma_{xy})_{\text{eff}}$. Setting $r^*=a$ gives a stress offset that now depends on the angle $\theta$ that the Burgers vector makes with the cylinder axis,
\begin{align*}
\left(\sigma_{xy}\right)^\dagger_{\text{eff}} &= sA/\cos(2\theta) = \sigma_{xy}^\dagger +2\pi A \gamma^{-1}  \tan(2\theta).
\end{align*}

Upon returning to the case of imposed axial tension, from Fig.~\ref{sigmadaggertheory}(a) we see that in order to have a plastic deformation event that decreases $R$, we need a reduced stress $\left(\sigma_{xx}-\sigma_{yy}\right)^{\dagger}_{ \text{eff}}/A\equiv \tilde \sigma_c(\phi)$ between $2$ and  $4/\sqrt{3}\approx 2.31$, depending on $\phi$. By symmetry, if $\left(\sigma_{xx}-\sigma_{yy}\right)_{\text{eff}}/A$ is \textit{negative} but greater in magnitude than the critical value $\tilde \sigma_c(\phi)$, then the tube will plastically deform by glide of dislocations that \textit{increase} $R$. The $+x$-moving dislocation will have Burgers vector opposite to that indicated in Fig.~\ref{sigmadaggertheory}(a), i.e., the directions coded \atwo, \maone, or \athree.   Eq.~(\ref{curvatureoffset}) therefore implies that the bending energy alone is  sufficient to destabilize narrow tubes with respect to dislocation unbinding events that increase $R$, provided $\gamma^{-1}$ is not too small. Specifically, for a given ratio $\kappa/Y$ and helical tilt $\phi$, there is a critical radius
\begin{align}
R_c &= \sqrt{\frac{4\pi \kappa}{\tilde \sigma_c(\phi) Y}} \label{Rcprediction},
\end{align}
below which a tube with $R<R_c$ will undergo a plastic deformation event that increases the radius spontaneously, even in the absence of any external stress. (Note that $\sqrt{\kappa/Y}$ must be at least of order $3a/2\pi $ in order for a tube with $R<R_c$  to be a geometrical possibility.)

The curvature corrections calculated above assume that $\gamma^{-1}$ is small. However, $\gamma^{-1}$ must still be large enough in order for this effect to be easily observable. \dbedit{Is this the case in single-walled carbon nanotubes?} SWCNTs have  unit cell spacing  $a\approx 0.24$ nm, a 2D Young's modulus of $Y\approx 340 $ J/m$^2$ \cite{yakobson1998mechanical,yao1998young,salvetat1999mechanical} and a  bending modulus calculated from monolayer graphene to be 	$\kappa\approx 2\times10^{-19}$ J \cite{lu2009elastic, wei2012bending}.  Thus, $\tilde \kappa= \kappa /Ya^2 \approx 0.01$, and $\gamma^{-1} = (a/R)^2 \tilde \kappa$ is in the range $10^{-3}-10^{-2}$ for typical carbon nanotubes with $2\pi R/a$ of order 10, suggesting that it would be difficult to measure the effects of the curvature energy in this system.\footnote{Thermal fluctuations are known to give rise to renormalized, scale-dependent bending rigidity $\kappa_R(\ell)$ and Young's modulus $Y_R(\ell)$ (where $\ell$ is the length scale), such that $\kappa_R(\ell)/Y_R(\ell)$ may be greater than $\kappa/Y$ by a factor $(\ell/\ell_{\mathrm{th}})^{\eta-\eta_u}$ \cite{2016PhRvB..93l5431K}. Here $\ell_{\mathrm{th}}$ is a thermal length-scale $\ell_{\mathrm{th}}\propto \sqrt{\kappa^2/(k_B T Y)}$ that is approximately 2 nm at room temperature, and $\eta-\eta_u \approx 0.46$. If the nanotube circumference $2\pi R$ is substituted for $\ell$, we find that thermal fluctuations give a renormalized reduced bending rigidity $\tilde \kappa_R(\ell) \equiv (\kappa_R(\ell)/Y_R(\ell)a^2)\approx c  \tilde \kappa   (R/a)^{0.46}$, where $\tilde \kappa =\kappa/Ya^2\approx 0.01$ and $c=(2\pi a/\ell_{\mathrm{th}})^{0.46}\approx 0.88$. Then the renormalized inverse F\"oppl-von K\'arm\'an number is $\gamma^{-1}_R(\ell)  = (a/R)^2  \times (\kappa_R(\ell) /Y_R(\ell) a^2) \approx 0.88 \tilde \kappa (R/a)^{-1.54}$ (still with $\ell =2\pi R$). The result is at most a very modest increase in $\gamma^{-1}_R(\ell) $ compared to the zero-temperature value $\gamma^{-1}= \tilde \kappa (R/a)^{-2}$. Raising the absolute temperature by a factor of 10 contributes merely a factor of $10^{0.46/2}\approx 1.7$ to $\gamma^{-1}$. Therefore, thermal effects are not expected to significantly change the results here.}  However, a curvature-induced force $\propto \kappa a/R_0^2$ on dislocation pairs has been noted in simulation studies of carbon nanotubes \cite{ding2007pseudoclimb}.  We note that molecular dynamics simulations of carbon nanotubes have found that, at least for zigzag tubes ($\phi=30^\circ$, corresponding to parastichy indices $m=n$ in our notation), the formation energy of a dislocation pair may become negative for $n<14$, although the radius of the $n=14$ zigzag tube is about an order of magnitude greater than $R_c\approx 0.2a$ \cite{PhysRevLett.81.4656}.


\section{Numerical modeling \label{secnum}}

To study how well our continuum elastic predictions apply to tubes of finite size, we now describe numerical simulations of dislocation glide in tubular crystals under axial stress $\sigma_{xx}>0$. The tubular crystal is modeled as a network of harmonic springs connecting nodes at the sites of a triangular lattice, following Ref.~\cite{seung1988defects}. The initial spring network gives the crystal a tubular topology and a particular choice of phyllotactic indices $(m,n)$, with each node connected to six springs so that the lattice is initially a pristine (defect-free) triangular crystal.

We implement periodic boundary conditions in the $X$ direction,  parallel to the cylinder axis $\bm{\hat x}$ (in the case of an exactly cylindrical tube), with springs joining nodes across the periodic boundaries. Periodic boundary conditions offer two advantages: We can ignore the complications of end-effects for dislocation dynamics, and the Gaussian curvature modulus does not contribute to the energy because the tube has no boundary.

As in Ref.~\cite{seung1988defects}, the stretching energy is discretized by defining a spring constant $\epsilon$ via a stretching energy
\begin{align}
F_s^{\text{discrete}} &= \tfrac{1}{2} \epsilon \sum_{\left<j,k\right>} \left(\left|\mathbf{R}_j - \mathbf{R}_k\right|-1\right)^2, \label{Fsdiscrete}
\end{align}
which corresponds to Eq.~\ref{Lame} with the choice $\mu=\lambda=\frac{\sqrt{3}}{4} \epsilon$. (Here, position vectors $\mathbf{R}$ are vectors in $\mathbb{R}^3$, and $\sum_{\left<j,k\right>}$ is a sum over neighboring nodes connected by an edge within a particular triangulation.) The Young's modulus and Poisson ratio are then $Y=\frac{2}{\sqrt{3}} \epsilon$ and $\nu=\frac{1}{3}$ respectively. Simulation lengths are scaled in units of the preferred lattice spacing, equal to the spring rest length. For a discrete version of the bending energy, we use the mean curvature energy discretization from Ref.~\cite{gompper1996random}, which can be written
\begin{align}
 F_b^{\text{discrete}} &= \kappa \sum_j \frac{ \left[ \sum_{k(j)}c_{jk} \left(\mathbf{R}_j-\mathbf{R}_k\right)\right]^2}{ \sum_{k(j)} c_{jk} \left(\mathbf{R}_j-\mathbf{R}_k\right)\cdot  \left(\mathbf{R}_j-\mathbf{R}_k\right)} \label{FbGK}
 \end{align}
where  $c_{jk} \equiv \cot \theta_{jk}^1 + \cot \theta_{jk}^2$, and $\theta_{jk}^{i}$ ($i=1,2$) are the angles opposite the edge $jk$ in the two faces to which that edge belongs.  $\sum_{k(j)}$ is a sum over neighboring nodes $k$ sharing an edge with node $j$. In the continuum limit, this bending energy corresponds to $F_b = \frac{1}{2} \kappa \int dx dy H^2,$ which for a cylinder of length $L$ and radius $R$ reduces to $\pi \kappa L/R$. An alternative curvature energy discretization is  \cite{seung1988defects} 
\begin{align}
F_b^\text{discrete,2} &= \frac{1}{\sqrt{3}}  \tilde \kappa \sum_{\left< \alpha,\beta \right>} \left|\mathbf{n}_\alpha - \mathbf{n}_\beta\right|^2 \nonumber \\
&=\frac{2}{\sqrt{3}} \tilde \kappa \sum_{\left< \alpha,\beta \right>} (1-\mathbf{n}_\alpha \cdot \mathbf{n}_\beta), \label{FbSN}
\end{align}
penalizing deviations between the unit normal directions of neighboring faces (summing over all pairs of neighboring faces). We do not use this form  because, even though Eq.~\ref{FbSN} has the correct continuum limit of $F_b = \frac{1}{2} \kappa \int dx dy \left(H^2-2K\right)$, it contains no information about stretching of the triangular faces, and therefore does not give the correct scaling $F_b \rightarrow  \pi \kappa L/R$ for pristine cylinders with periodic boundary conditions under external forces. In contrast,we have checked that  the discretization in Eq.~\ref{FbGK} for the bending energy converges to $\pi \kappa L/R$ with $1\%$ accuracy for a wide variety of \dbedit{strained and unstrained} tubular crystals, including those as small as $R\approx 3a$.

\begin{figure}[htbp]
\includegraphics[width=\linewidth]{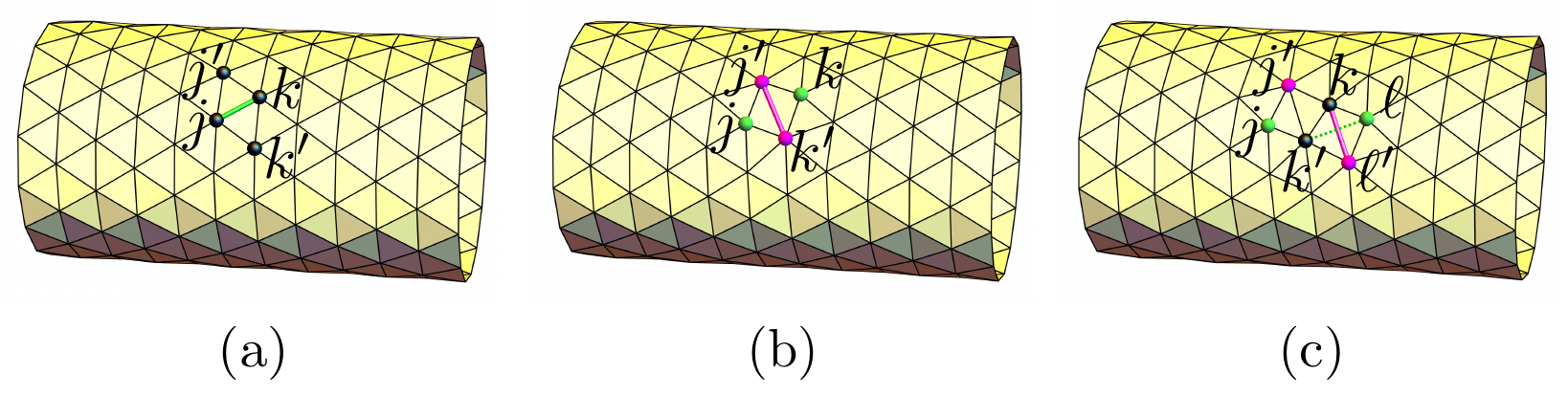}
\caption{\dbedit{Illustration of dislocation nucleation and glide by bond flips. (a) In a pristine lattice, with all nodes six-coordinated, a bond $jk$ between nodes $j$ and $k$ (shown in green) is chosen to be flipped. (b) Bond $jk$ is replaced by a new bond $j'k'$ (magenta) connecting the common neighbors of $j$ and $k$. This nucleates two dislocations $\{j,j'\}$ and $\{k,k'\}$. Five-coordinated and seven-coordinated disclinations are labeled by green and magenta spheres, respectively. (c) Another bond flip, replacing bond $k'\ell$ (green dashed line) with a new bond $k\ell'$ (magenta), glides the rightmost dislocation from $\{k,k'\}$ to $\{\ell,\ell'\}$; nodes $k$ and $k'$ are now once again six-coordinated. } \label{bondflipfig}}
\end{figure}

Recall that the lattice sites of a tubular crystal are well-approximated by the centers of discs packed in a cylindrical surface. Therefore, we use the latter geometry as an initial state and then minimize the energy over node positions to obtain the preferred lattice sites of the pristine tubular crystal. A dislocation pair is then created from the pristine lattice by a bond flip that removes a bond between neighboring nodes $j,k$ and replaces it with a new bond between nodes $j'$ and $k'$, the common neighbors of $j$ and $k$, as shown in Fig.~\ref{bondflipfig}(a,b). The node pairs $\{j,j'\}$ and $\{k,k'\}$ are then a pair of dislocations with equal and opposite Burgers vectors, each comprised of a positive disclination at a five-coordinated node ($j$ or $k$) and a negative disclination at a seven-coordinated node ($j'$ or $k'$). Thereafter, dislocation glide of $\{k,k'\}$ by one lattice spacing to a neighboring node pair $\{\ell, \ell'\}$ is accomplished by a similar bond flip, removing the bond $k'\ell$ and replacing it with a new bond $k\ell'$, as depicted in  Fig.~\ref{bondflipfig}(c). By repeating this process on either or both dislocations, we can adjust the glide separation distance $r$ of the dislocation pair in discrete steps, always assuming a small but finite temperature ensures enough energy to easily surmount the small Peierls potential. 

For each spring network, the  total energy $F_\text{tot}=F_s^\text{discrete}+ F_b^\text{discrete}$ is minimized over the positions $\{\mathbf{R}_i\}$ of all nodes using a conjugate gradient algorithm from the ALGLIB package \cite{ALGLIB}.  We probe the energy landscape of dislocation glide by comparing $F_\text{tot}$ for a given dislocation glide separation to $F_\text{tot}$ with the dislocations one glide step closer together or one step farther apart. The option giving the lowest $F_\text{tot}$ determines the new state of the system. There are three possible final outcomes: The dislocations may annihilate into the defect-free state; they may reach a separation along the cylinder axis of at least half the length of the perfect cylinder; or they may come to rest at some smaller but nonzero separation. In the case of the second outcome, we consider the dislocations to be ``free'', i.e., on a truly infinite cylinder we expect they would continue gliding to infinite separation. If the dislocations annihilate or come to rest at a smaller separation, we consider them to be still bound. 

Periodic boundary conditions in $X$ require two more independent variables in addition to the node positions: the horizontal length $L_X$ of the box, and a rotation angle $\beta$ by which the right end of the cylinder is rotated about the $\bm{\hat X}$ axis before being reconnected with the left end. Chiral lattices generically necessitate nonzero values of $\beta$. We minimize over $L_X$ and $\beta$ simultaneously with the node positions $\{\mathbf{R}_i\}$. Before the application of any strain, we record the ``rest'' values $L_X^0$ and $\beta^0$ of these variables. Then, we can apply an axial strain $u_{xx}$ by holding $L_X$ fixed at $L_X^0(1+u_{xx})$. A torsional strain could be applied by holding $\beta$ fixed at $\beta^0\left(1-(L/R)u_{xy}\right)$ \cite{zhang2009dislocation}. 

\begin{figure}[htbp]
\includegraphics[width=\linewidth]{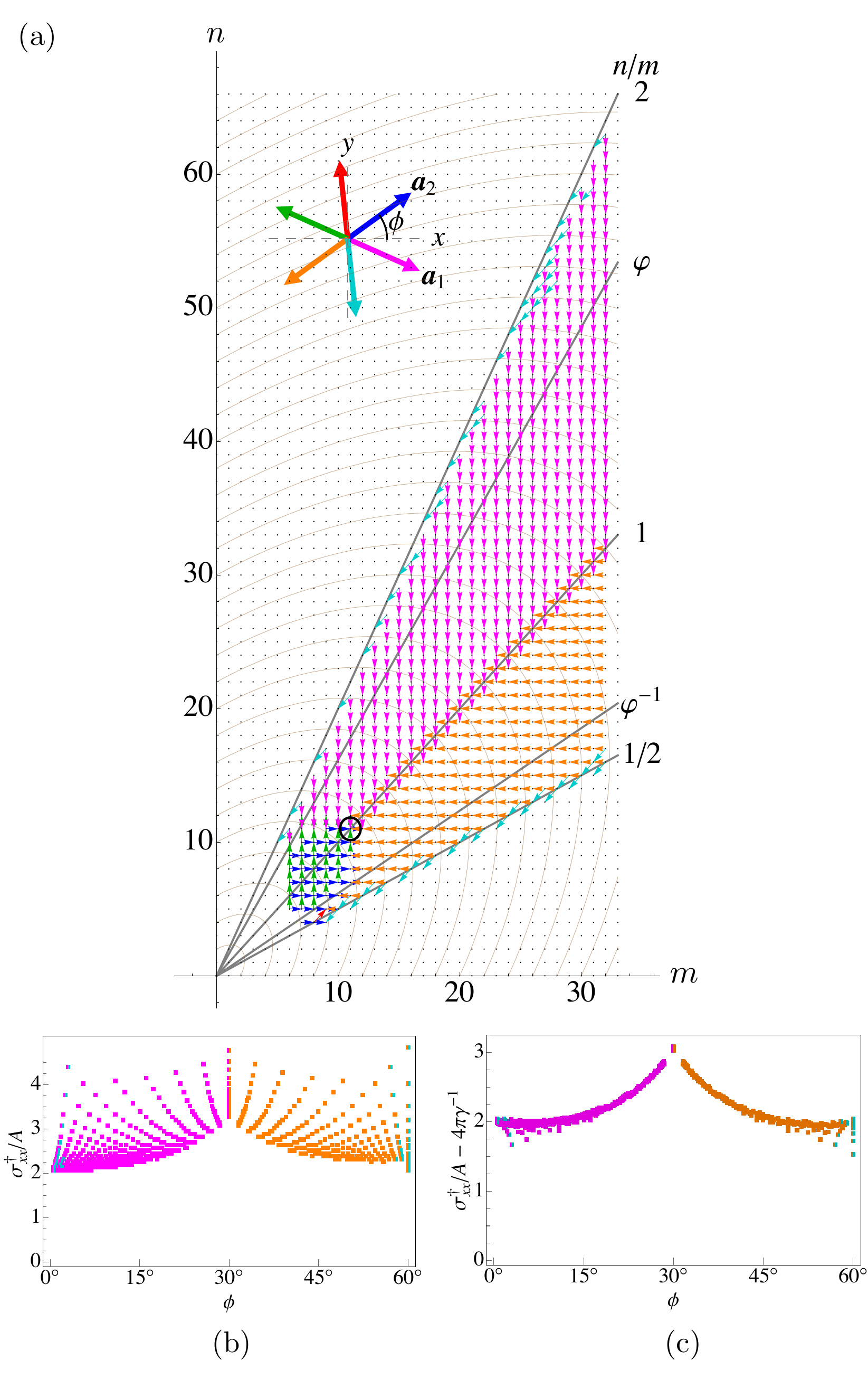}
\caption{Numerical results for plastic deformation of tubular crystals (accompanied by changes in the parastichy numbers $(m,n)$) under axial tension $\sigma_{xx}>0$ with a reduced bending modulus  $\tilde \kappa =\kappa/Ya^2 = 0.5$. The dislocations that unbind at lowest applied stress (consistent with the allowed transitions summarized in Table~\ref{bmntable}) are recorded, using the color scheme as in shown in the inset to (a). (a) Arrows connecting grid points indicate the transformation of the tubular crystal from $(m,n)$  to $(m',n')$. Brown elliptical contours are curves of constant $R\propto \sqrt{m^2+n^2-mn}$. Lines of constant $\phi$ (i.e., constant $n/m$) are shown in gray, with  $n/m$ as marked on the right side. Here, $n/m=1$ corresponds to $\phi =30^\circ$. (For comparison, we show the golden mean $\varphi =\frac{1}{2}(1+\sqrt{5})$ and its inverse, favored in many instances of plant phyllotaxis \cite{adler1997history,kuhlemeier2007phyllotaxis}.) Blue, \maone, and \athree\  arrows are tube-widening events triggered by the bending rigidity that occur at zero applied stress.  The point ringed by a black circle marks the widest tube with $m=n$ that is unstable to spontaneous tube-widening dislocation pairs, driven by the bending energy. This gives the critical tube radius $R_c/a\approx 11/(2\pi)$ for $\phi=30^\circ$ and $\tilde\kappa=0.5$, corresponding  to the data point ringed by a black circle in Fig.~\ref{Rcvskappa}.  (b) The critical applied axial stress $\sigma_{xx}^\dagger$ required for each event in (a) as a function of $\phi$, with strings of closely spaced squares swept out as $R$ varies.  (c) The same data as in (b) but collapsed by including the curvature correction from Eq.~(\ref{curvatureoffset}). \label{sigmaxxnum}}

\end{figure}

The numerical results for plastic deformation under axial tension are plotted in Fig.~\ref{sigmaxxnum} for a range of $(m,n)$ initial tubular crystal tessellations. Starting from zero, an applied axial strain is slowly increased in steps of $1\%$. At small strains, any dislocation pair annihilates immediately after it is nucleated, returning to the pristine lattice. At a critical strain $\sigma_{xx}^\dagger$,  a dislocation pair with one of the six available Burgers vector pairs unbinds and glides apart to freedom. (For some $(m,n)$ values, two Burgers vectors unbound at the same $\sigma_{xx}^\dagger$.) Fig.~\ref{sigmaxxnum}(a) records with colored arrows the plastic deformations $(m,n)\rightarrow(m',n')$ obtained in this way, using the Burgers vector coloring scheme introduced in Fig.~\ref{introfig}. These arrows collectively indicate the flow through the tessellation parameter space of $(m,n)$ values (or, equivalently, quantized $(R,\phi)$ values) under axial tension. The overall response to an axial stress $\sigma_{xx}>0$  is of course a step-by-step decrease in $R$.\footnote{Negative $\sigma_{xx}$ or a torsion $\sigma_{xy}$ have the additional complications of buckling or supercoiling, instabilities which are outside the scope of this paper.} Meanwhile, there is a convergence of the lattice orientation toward the $m=n$ line of achiral states, where $\phi=30^\circ$, as predicted in Fig.~\ref{sigmadaggertheory}(a). Because of the stepwise nature of the plastic deformation toward smaller $R$, with $(\Delta m,\Delta n)=(0,-1)$ or $(-1,0)$ at each step, the parastichy tilt angle angle $\phi$ oscillates slightly about $30^\circ$ as the tube radius shrinks, as also found for carbon nanotubes \cite{yakobson1998mechanical}. \dbedit{Plastic deformations with $(\Delta m,\Delta n)=(-1,-1)$, with $\mathbf{ b} = - \mathbf{a}_3$ (coded cyan), were recorded only near $\phi = 0$ or $60^\circ$, where Fig.~\ref{sigmadaggertheory}(a) shows that   the theoretically predicted $\left(\sigma_{xx}-\sigma_{yy}\right)^\dagger$ for  $\mathbf{b}=-\mathbf{a}_3$ coincides with the minimal $\left(\sigma_{xx}-\sigma_{yy}\right)^\dagger$  associated with $\mathbf{ a}_1$ or $-\mathbf{a}_2$. }

The critical axial stress $\sigma_{xx}^\dagger$ required to pull apart the dislocations recorded in Fig.~\ref{sigmaxxnum}(a) is plotted in Fig.~\ref{sigmaxxnum}(b) as a function of $\phi$. In Fig.~\ref{sigmaxxnum}(c), the data collapses to a single curve when the curvature offset to $\sigma_{xx}$ due to the final bending rigidity term in Eq.~(\ref{curvatureoffset}) is included. The scatter in Fig.~\ref{sigmaxxnum}(b) reflects the $R$-dependence of $\sigma_{xx}^\dagger$, but Fig.~\ref{sigmaxxnum}(c) confirms that the $R$-dependence is described simply by the curvature offset $-Y \gamma^{-1} =-\kappa /R_0^2$, where $R_0$ is the radius that minimizes the stretching energy $F_s$. With this correction, the critical stress depends only on $\phi$. While  Fig.~\ref{sigmaxxnum}(a) shows results for reduced bending modulus $\tilde \kappa\equiv \kappa/Ya^2 = 0.25$, similar results were obtained for $\tilde \kappa = 0.5$, $0.75$, and $1$.   The Burgers vectors and the magnitude of $\sigma_{xx,\text{eff}}^{\dagger}$ in Fig.~\ref{sigmaxxnum}(c) are in approximate agreement with the predictions of Fig.~\ref{sigmadaggertheory}(a). However, the exact shape of the $\sigma_{xx,\text{eff}}^{\dagger}$ curve as a function of $\phi$ differs somewhat between theory and numerics, reflecting differences between the continuum and discrete formulations.

\begin{figure} [htbp]
\centering \includegraphics[width=0.8\linewidth]{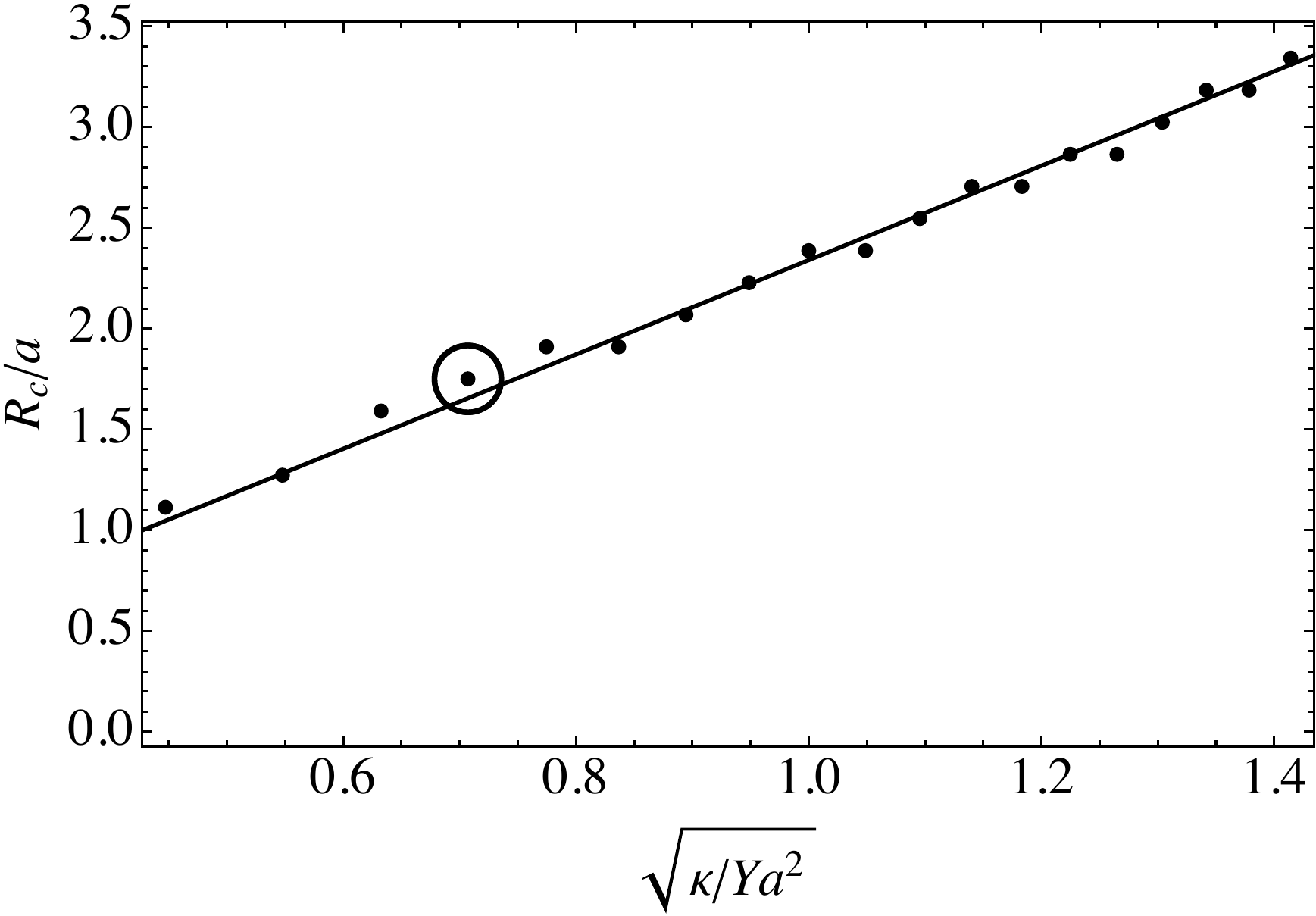}
\caption{Critical tube radius $R_c$, below which bending energy causes spontaneous dislocation unbindings that increase the tube radius (for $m=n$ achiral tubes only), as a function of the curvature modulus $\kappa$. The curve is a best fit $R_c = \sqrt{1.74 \pi \kappa/Y}$, in good agreement with the prediction $R_c \approx \sqrt{1.73\pi \kappa/Y}$ from Eq.~(\ref{Rcprediction}) (with $\phi=30^\circ$). The data point ringed by a black circle corresponds to the circled $(m,n)=(11,11)$ grid point in Fig.~\ref{sigmaxxnum}(a). \label{Rcvskappa}}
\hrule 
\end{figure}

An interesting feature of Fig.~\ref{sigmaxxnum}(a) is the presence of \maone, \atwo, and \athree\ arrows at small $R$, indicating plastic deformation events that increase $R$, contrary to the predominant  \matwo, \aone, and \mathree\ arrows that all show the tube radius shrinking. These tube-widening deformations occur at zero applied axial strain, and arise solely from the effective stresses $\sigma^b_{xx}$ and $\sigma^b_{yy}$ generated by the bending energy as given in Eq.~\ref{sigmab}. As predicted, there is a critical radius $R_c$ below which the bending energy alone makes narrow tubes unstable to widening. To compare quantitatively with theory, we fixed $\phi=30^\circ$ and varied $m=n\approx2\pi R/a$ to find the largest radius $R_c$ at which a tube-widening deformation occurs spontaneously. This procedure was repeated over a range of values of the reduced bending modulus $\tilde \kappa\equiv \kappa/Ya^2$. The results are plotted in Fig.~\ref{Rcvskappa}, which shows that $R_c/a$ varies as $\sqrt{\tilde \kappa}$. A linear fit gives $R_c =\sqrt{1.74 \pi \kappa /Y}$, which agrees well with the the prediction in Eq.~(\ref{Rcprediction}), $R_c \approx \sqrt{1.73\pi \kappa/Y}$, using $\tilde \sigma_c (\phi=30^\circ)=4/\sqrt{3}$. 

\begin{figure}[h!]
\includegraphics[width=\linewidth]{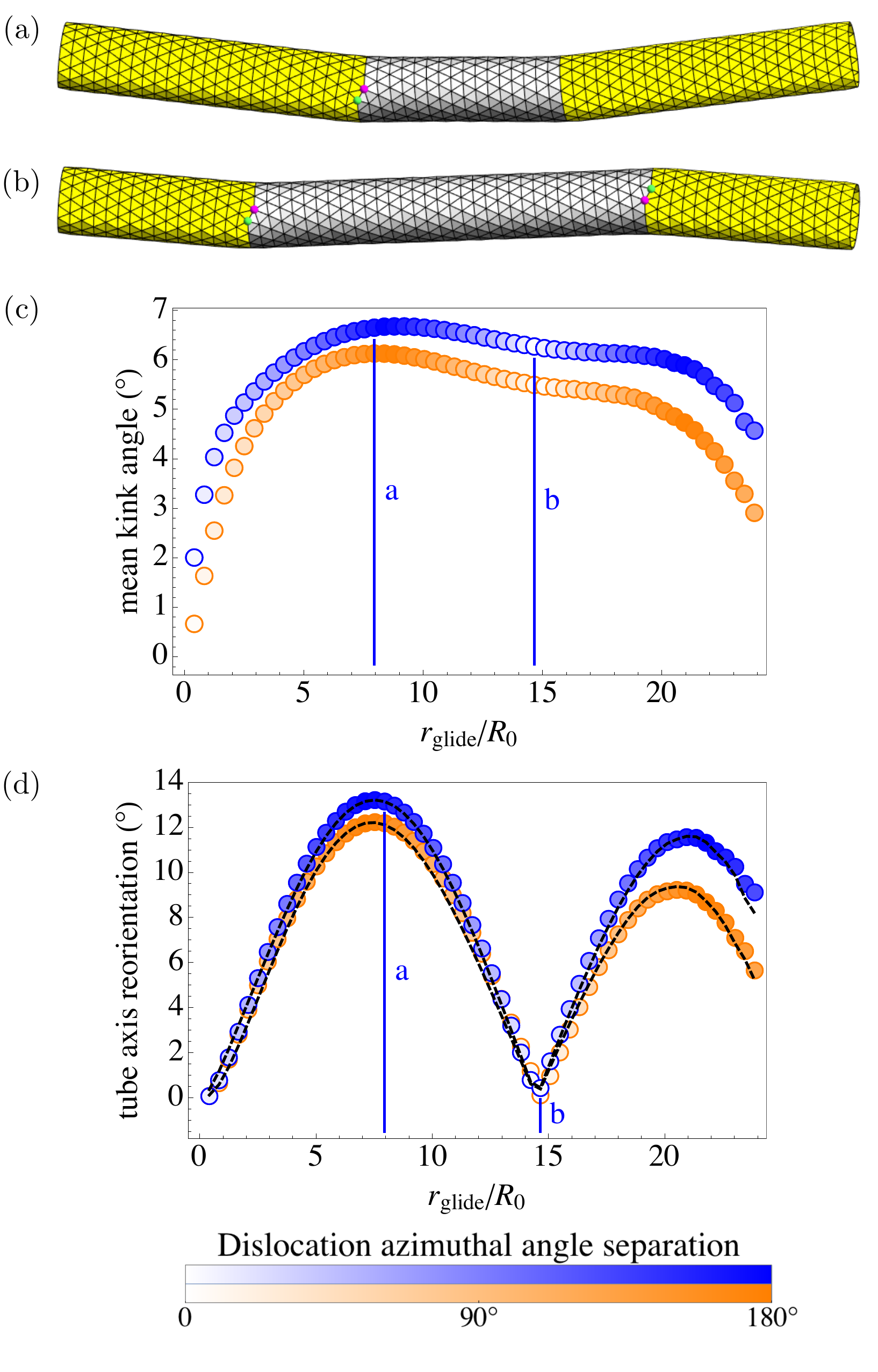}
\caption{(a,b) Two snapshots from the glide process of a simulated tubular crystal undergoing a plastic deformation from $(m,n)=(15,15)$ (yellow) to $(m',n')=(15,14) $ (gray), with reduced bending rigidity $\kappa /Ya^2 \equiv \tilde \kappa=0.09$. Green and magenta spheres indicate the dislocations' five- and seven-coordinated disclinations, respectively. (The rightmost dislocation is on the far side of the tube in (a).)  The external stress is set to zero at each snapshot. (c) Plot of the tube axis kink angle, averaged over the two dislocation sites, as a function of the dislocations' glide separation distance in units of the initial tube radius $R_0$. (d) The angle between the leftmost and rightmost segments of the tube axis, measuring reorientation of the tube axis by the $(m',n')$ region, as a function of dislocation glide separation distance. Dashed black curves are the geometric model for tube axis reorientation as given in Eq~\ref{betaeqn}. In both (c) and (d): Blue and orange data points correspond to reduced bending modulus $\tilde\kappa=0.09$, $0.9$, respectively; color saturation of data points indicates the dislocations' azimuthal angle separation as shown in the legend at bottom; and vertical blue lines correspond to snapshots (a) and (b) as indicated.\label{kinkfig}} 
\end{figure}

The simulated tubes reveal that, in the presence of dislocations, the deviations from a perfect cylindrical shape are not limited to changes in radius: In addition, the dislocations create  kinks in the tube axis, as can be seen in Fig.~\ref{kinkfig}(a,b). As pointed out in  Ref.~\cite{amir2013theory}, a pair of dislocations in a crystal on a perfectly cylindrical surface act like a pair of grain boundaries, between which the crystal axes are slightly reoriented. We find that when the tubular crystal is free to assume an energy-minimizing shape, it is approximately piecewise cylindrical (far from the necks discussed in the following section) with a kink angle in the tube axis associated with a reorientation of the crystal axes across the boundaries between the different $(m,n)$ tessellations. Thus while one pair of dislocations is gliding apart, the tube axis contains two kinks at the boundaries between the outer $(m,n)$ and inner $(m',n')$ tessellations. (Here we approximate the tube axis using a computed ``spine curve'' as described in the next section. We also remove the periodic boundary conditions along the tube axis to investigate bending of tubes by dislocations. After each glide step, the applied strain is set to zero and the tube conformation is relaxed with the dislocations frozen in place.)

Averaging over the two tube axis kinks, we find a kink  angle on the order of $a/|\mathbf{C}|$, or (equivalently) of the difference $\Delta \phi$ between the helical orientations of the two tessellations, provided that the dislocation separation $r_{\text{glide}}$ exceeds the circumference $\approx 2\pi R_0$. This average kink angle is plotted in Fig.~\ref{kinkfig}(c) as a function of  $r_{\text{glide}}$ for a $\mathbf {b} = \pm \mathbf{a}_1$  dislocation pair  during the plastic deformation process $(m,n)=(15,15)\rightarrow(m',n')=(15,14)$. Because the dislocations move along their helical glide paths, the azimuthal angle separation of a defect pair (shown as color saturation of data points in Fig.~\ref{kinkfig}(c,d)) varies linearly with $r_{\text{glide}}$. Blue and orange data points correspond respectively to reduced bending modulus $\tilde \kappa = 0.09$, $0.9$, showing that there is only a weak dependence of this effect on $\tilde \kappa$. Figs.~\ref{kinkfig}(a,b) are snapshots from the $\tilde \kappa=0.09$ case. 

 The net effect of the pair of kinks on the tube axis orientation depends crucially on the relative azimuthal coordinates of the two dislocations. If the two dislocations are on opposite sides of the tube, as in Fig.~\ref{kinkfig}(a), their respective kink angles add constructively, effectively bending the tube. In contrast, if the two dislocations are on the same side of the tube, as in Fig.~\ref{kinkfig}(b), the two kinks effectively cancel each others' reorientations of the tube axis, producing a tube conformation that is not bent but instead zigzagged. Fig.~\ref{kinkfig}(d) shows the angle between the tube axis in the left $(m,n)$ region and the tube axis in the right $(m,n)$ region, i.e., the change in tube axis orientation after traversing the $(m',n')$ region. The data is again plotted as a function of dislocation pair glide separation $r_{\text{glide}}$ and shows only a weak dependence on $\tilde \kappa $. The tube axis reorientation angle is near its maximum in Fig.~\ref{kinkfig}(a) and nearly zero in Fig.~\ref{kinkfig}(b), with intermediate azimuthal separation angles of the defects producing states intermediate between the bent and zigzagging tube conformations. 

We can understand the shapes of the tube reorientation angle curves in Fig.~\ref{kinkfig}(d), producing the dashed black-line approximations shown there, through the following geometric reasoning. Assume that the tube axis $\bm{\hat t}_C$ in the central region is along $\bm{\hat X}$, and that each of the two dislocations causes a deflection in the tube axis by a small angle $\delta$ in the direction $\bm{\hat X}\times \bm{\hat \rho}$. Here, $\bm{\hat{\rho}}$ is the unit radial vector pointing from the tube axis to the dislocation (see Fig.~\ref{zetaschemfig}). Then the tube axis on the left side of the tube has unit tangent vector $\bm{\hat t}_L \propto \bm{\hat{X}}+\delta (\bm{\hat{X}}\times \bm{\hat{\rho}}_L)$, and likewise on the right side of the tube $\bm{\hat t}_R \propto \bm{\hat{X}}+\delta (\bm{\hat{X}}\times \bm{\hat{\rho}}_R)$, where $\bm{\hat{\rho}}_L$ and $\bm{\hat{\rho}}_R$ correspond respectively to the left and right dislocations. For a dislocation at azimuthal angle coordinate $\alpha$ relative to $\bm{\hat Y}$, we have $\bm{\hat X}\times \bm{\hat \rho} = \bm{\hat Z} \cos \alpha - \bm {\hat Y} \sin \alpha$. After normalizing $\bm{\hat t}_L$ and $\bm{\hat t}_R$, it is straightforward to determine that
\begin{align}
\bm{\hat t}_L \cdot \bm{\hat t}_R &= \frac{1+\delta^2 \cos (\Delta \alpha )}{1+\delta^2} \label{tldottr}
\end{align}
where $\Delta \alpha$ is the difference in azimuthal angular coordinate $\alpha$ between the dislocations. Setting Eq.~(\ref{tldottr}) equal to $\cos \beta$ where $\beta$ is the (small) tube axis reorientation angle, we find
\begin{align}
\beta &\approx \delta \sqrt{2\left(1-\cos(\Delta \alpha)\right)} \label{betaeqn}.
\end{align} 
In Fig.~\ref{kinkfig}(d), at each data point we use for the deflection angle $\delta$ the numerically measured mean kink angle from Fig.~\ref{kinkfig}(c). The dislocations' azimuthal separation $\Delta \alpha$ increases linearly with glide separation $r_{\text{glide}}$, with slope determined by the helical pitch of the associated parastichy. The tube axis reorientation angle $\beta$ is then calculated using Eq.~\ref{betaeqn}, producing the dashed black lines in good agreement with the reorientation angle measured directly from the simulated tubes.

\section{Tube necks \label{necksec}}

So far our analytic calculations have assumed the tube has a perfect cylindrical shape, \dbedit{albeit with a radius that can change in time in response to external stress}. However, at any intermediate stage of the dislocation pair's glide separation, the radius necessarily changes \textit{spatially} at a pair of necks in the tube, where the two dislocations interpolate between regions with different preferred circumferences $|\mathbf C(m,n)|$.  Given the stepwise nature of the plastic deformation mechanism, the fractional change in radius mediated by  each dislocation is typically small, so the cylindrical shape is a reasonable starting approximation. On the other hand, we can use the fractional change in radius as a small parameter, allowing some analytic insight into the equilibrium shape of a tubular crystal interrupted by dislocations, as discussed below. (How these small  deviations  modify the energy landscape of dislocation motion during plastic deformations is a more subtle problem that we leave for future work.) 

In this section, we first apply a local measure of the tubular crystal's radius to compute the profile of the tube's neck in the numerical simulations. We then compare these results with an analytic treatment of  small deflections in thin-walled cylinders.

\begin{figure}[htbp]
\includegraphics[width=\linewidth]{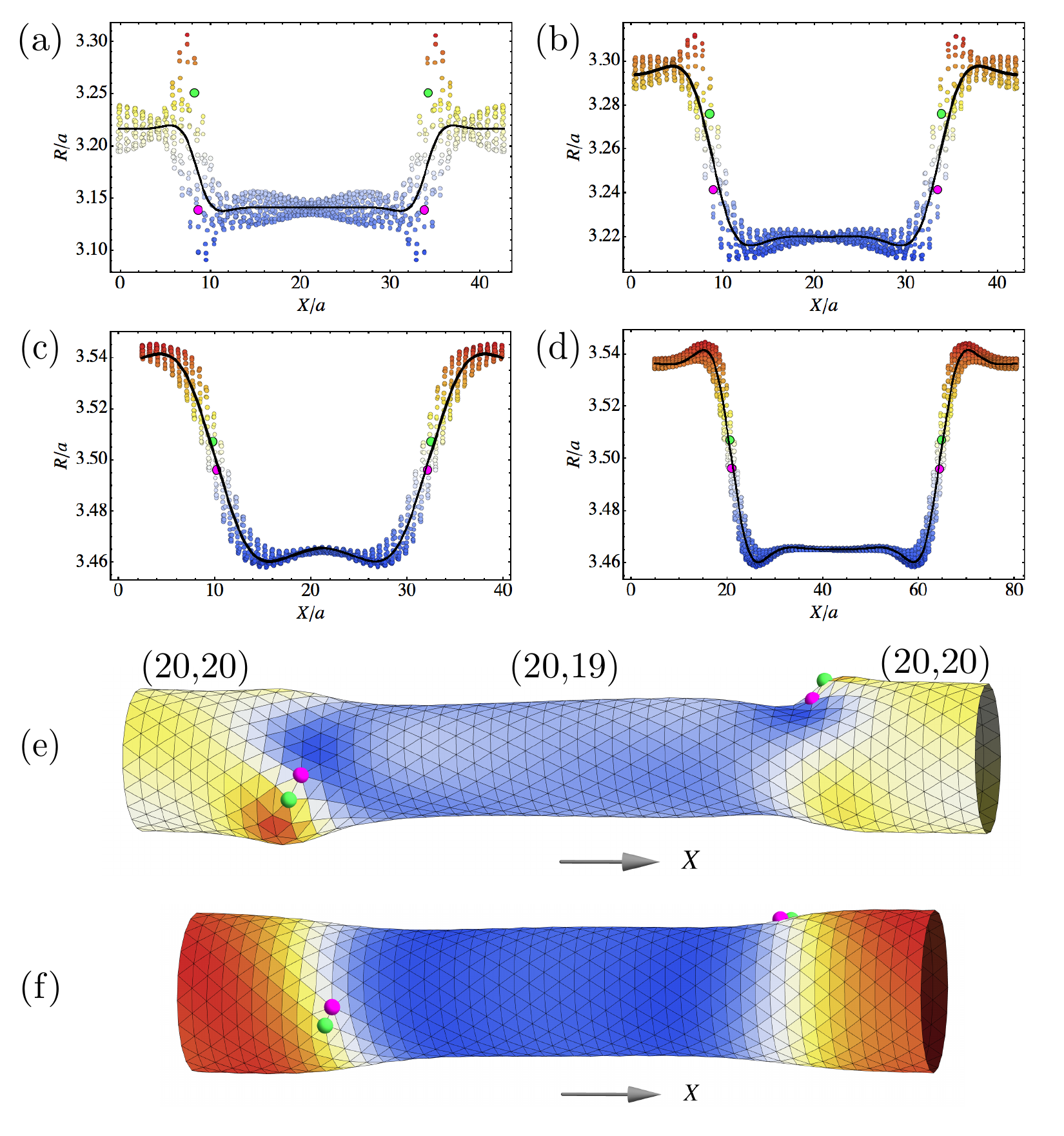}
\caption{ (a-d) Tube radius as a function of coordinate $X$ along the  axis of the unperturbed cylinder (both in units of $a$), as a dislocation pair whose right-moving dislocation has Burgers vector $\mathbf b = \mathbf a_1$ interpolates between  an $(m,n)=(20,20)$ tessellation on the left and right sides and an $(m,n)=(20,19)$ tessellation in the middle. Green and magenta dots mark $X$-coordinates of the 5- and 7-coordinated disclinations; colors of other data points are a proxy for the local tube radius $R$. The reduced bending rigidity $\tilde \kappa = \kappa/Ya^2$ is   0.1 (a),  0.5 (b), 2 (c-d), while the number of lattice sites is $N=1000$ (a-c) or $N=2000$ (d). The spread in radii at a given $X/a$ is due to the azimuthal variation in the distance to the spine. Theoretical curves are from Eqs.~(\ref{zetaplus})-(\ref{cplus}). (e) A heat map of the local radius on the tubular crystal corresponding to (a), with the colors corresponding to the $R$ values as shown there. (f) A similar heat map of the local radius corresponding to (c). In both (e) and (f), radius variations in the positions of plotted surface points, relative to an average-radius reference cylinder, have been exaggerated by a factor of 10 for clarity.\label{Rvsglide} }

\end{figure}

Defining a  local radius for the simulated tube is somewhat subtle. For the tubes shown in Fig.~\ref{Rvsglide}(e,f), isolated dislocations cause slight bending or reorientation of the tube axis (as we have seen in the previous section), which complicates the determination of a centerline from which to measure the radius on the tube surface. To define a tube centerline, or ``spine'', we adopt the following method: We fix $X$ (the coordinate along the unperturbed cylinder's axis), and then we find the $Y$ and $Z$ coordinates (in $\mathbb{R}^3$)  of a point on the spine by finding the zero of a fictitious repulsive force exerted by every lattice site in the tubular crystal on the spine point (with the $X$ component of the force projected out). We found it convenient to impose a  fictitious force pointing along the separation direction $\mathbf d$ between a tube lattice site and the spine point, and varying like $1/|\mathbf d|^8$. In tests on point sets sampled from analytically defined canal surfaces \cite{HilbertGeometry}, this procedure  produced a computed spine in good agreement with the known spine curves of these continuous surfaces.   Linear interpolation was then used to fill in the spine after a certain number of points had been calculated in this way. The local radius at a particular tube lattice site was then estimated as the distance from that site to the nearest point on the calculated spine.

Computed local radius results are shown in Fig.~\ref{Rvsglide} for an intermediate stage in the plastic deformation  $(m,n)=(20,20) \rightarrow (20,19)$ by a dislocation pair whose right-moving dislocation has Burgers vector $\mathbf b = \mathbf a_1$. (Here the axial strain is set to zero before measuring the radius profile.) Three different values of the reduced bending modulus $\tilde \kappa=\kappa/Ya^2$ are used in Fig.~\ref{Rvsglide}(a-d). $R$ is plotted against $X$ (the coordinate in $\mathbb{R}^3$ most closely aligned with the cylinder axis), suppressing the azimuthal coordinate. For the largest tested bending modulus $\tilde \kappa =\kappa/Ya^2= 2$ in Fig.~\ref{Rvsglide}(c,d), $R$ is approximately independent of the azimuthal coordinate, whereas for small $\tilde \kappa$ there is considerable azimuthal variation in $R$. In Fig.~\ref{Rvsglide}(a), where $\tilde \kappa = 0.1$,  the disclination pairs comprising the dislocations create sharp local disturbances in $R$, with the positive disclination at a higher $R$ and the negative disclination at a lower $R$ than the radius of either the initial or final pristine tubes. In contrast, at larger $\tilde \kappa$ the disclinations do not generate a large local change in $R$, instead following the gentler slope of the neck (Fig.~\ref{Rvsglide}(c,d)). The transition between these two extremes is gradual, as illustrated in Fig.~\ref{Rvsglide}(b) for $\tilde \kappa = 0.5$.

The large jump in $R$ between the disclinations at low $\tilde \kappa$ in Fig.~\ref{Rvsglide}(a) can be understood in light of the known buckling behavior of a membrane with a dislocation. In Ref.~\cite{seung1988defects}, a membrane with a dislocation in its center and lattice spacing $a$ was found to buckle, with a sudden jump in the out-of-plane direction between the two disclinations, when the membrane's linear size exceeded $ (127 \pm 10) a \tilde \kappa $. Upon setting this linear size equal to $2\pi R_0$, we find a critical value $\tilde \kappa_b \approx 2\pi R_0 / (127 a)$ below which buckling is expected. For the $(m,n)=(20,20)$ tube, $\tilde \kappa_b \approx 0.16$, in agreement with the buckled behavior we see with $\tilde \kappa = 0.1$.

We now seek  analytic insight into the shape of the neck in the opposite limit where $\tilde \kappa \gg \tilde \kappa_b$ and there is a well-defined ``neck profile'' $R(X)$. In particular, we would like to understand why $R(X)$ in Figs.~\ref{Rvsglide}(b-d) exhibits oscillatory behavior in addition to an overall rise or decay between the narrower and wider region of the tube. 

\begin{figure}[htbp]
\centering
\includegraphics[width=0.8\linewidth]{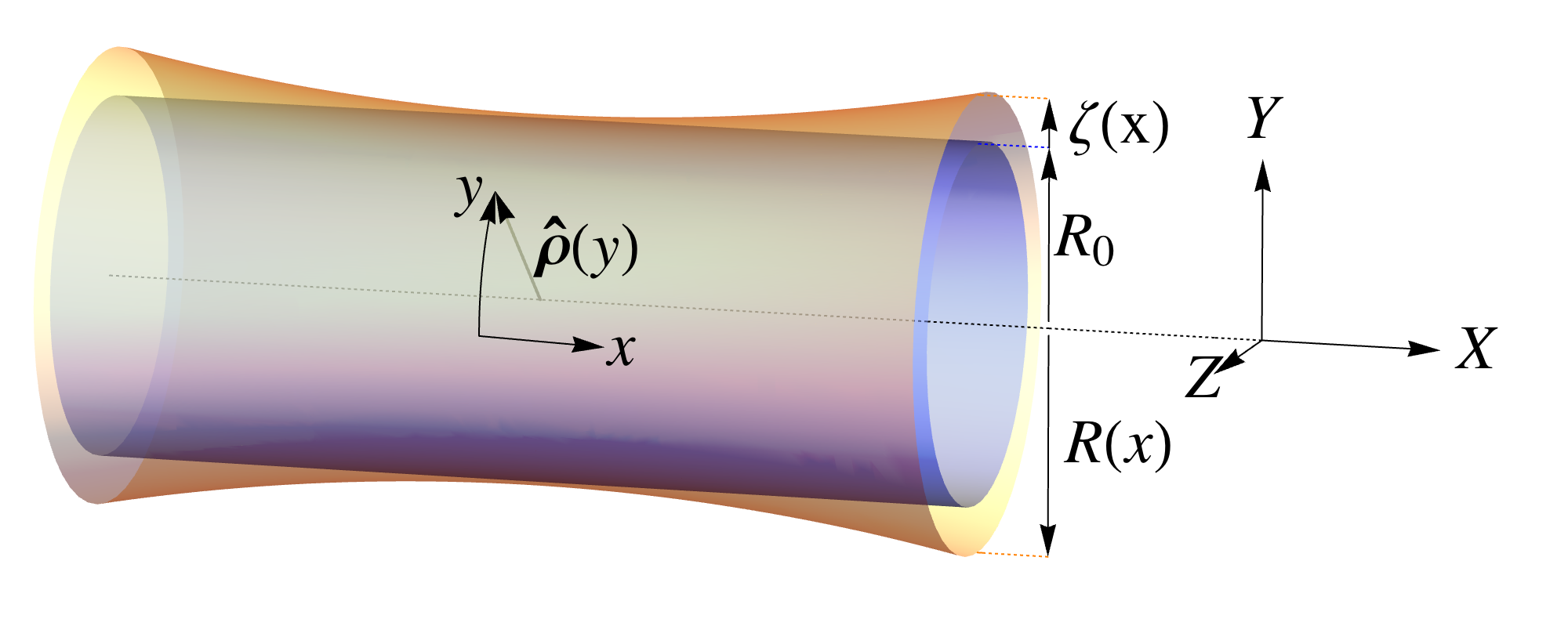}
\caption{Schematic illustration of the coordinate systems and variables used in Sec.~\ref{necksec} to describe a weakly distorted cylinder. An ideal  cylinder (blue) of radius $R_0$ is transformed into a surface of revolution (orange) with azimuthally-symmetric but $x$-dependent radius $R(x) = R_0+\zeta(x)$. Lowercase $(x,y)$ are coordinates in the undeflected cylindrical reference surface, respectively along the axial and azimuthal directions. Uppercase $(X,Y,Z)$ are coordinates of the three-dimensional embedding, with the tube axis (dotted line) oriented along $\bm{\hat X}$. The $\bm{\hat \rho}(y)$ direction points radially out from the tube axis in three dimensions. \label{zetaschemfig}}
\end{figure}

Consider weak radial deflections from a perfect cylinder of radius $R_0$, described by a function $\mathbf{r}(x,y)$ in $\mathbb{R}^3$:
\begin{align}
 \mathbf{r}(x,y) &= \left[ R_0+   \zeta(x,y) \right] \bm{\hat \rho}(y) + x\bm{ \hat X },
\end{align}
where we define a reduced tube deflection $\tilde \zeta(x,y) \equiv \zeta(x,y)/ R_0 \ll 1$ and $\bm{\hat \rho}(y)$ is the unit radial vector $\bm{\hat Y} \cos(y/R_0) +\bm{ \hat Z} \sin(y/R_0)$. These coordinates and variables are illustrated in Fig.~\ref{zetaschemfig}. 
The mean curvature for this parametrization is
\begin{align}
H &\approx \zeta_{xx} + \zeta_{yy} - R_0^{-1}(1-\zeta R_0^{-1})  - 2 \zeta \zeta_{yy} R_0 ^{-1} ,
\end{align}
where subscript $x,y$ denote spatial derivatives.
As seen in Fig.~\ref{Rvsglide}(c,d), for larger $\tilde \kappa$ the radius is approximately independent of the azimuthal coordinate. In this regime we can therefore make the simplifying assumption that  $\zeta$ has no dependence on $y$. Then

\begin{align}
H &= \zeta_{xx} - R_0^{-1}(1-\zeta R_0^{-1}) .\label{Happrox}
\end{align}

We can  estimate  the width $w$ of the neck along the cylinder axis using scaling arguments:  Assume that over the neck  width $w$, there is  a change in  radius $\Delta \zeta$ of order $a$, since $R_0\approx (a/2\pi)\sqrt{m^2+n^2-mn} $ and  each dislocation changes $m$ and/or $n$ by at most $\pm 1$. The stretching energy density is then $ \sim Y (a/R)^2 $. 
\dbedit{The leading-order term in the squared  mean curvature  is $\zeta_{xx}^2\sim (a/w^2)^2$, giving a curvature energy density $\sim \kappa(a/w^2)^2 $. (Other terms in $H^2$ are smaller by factors of $w/R$, which is small, as shown below.)}  Equating the bending and stretching energy density scalings gives $w \sim \gamma^{-1/4} R$, similar to deformations on a sphere \cite{paulose2012fluctuating}. Thus, $w/R$ scales like the inverse fourth root of the F\"oppl-von K\'arm\'an number, which we assume is large. (See also Ref.~\cite{landau1975elasticity}, where a point force $f$ is applied normal to a thin shell of thickness $h$, and the shell is deformed by a height $\zeta \sim fR/Yh$, over a region of size $d\sim \sqrt{h R} \sim \gamma^{-1/4} R$.) In contrast, large-amplitude ``pinch in a pipe''  studied by Mahadevan \textit{et al.} \cite{mahadevan2007persistence} has a persistence length $l_p \sim \gamma^{1/4} \sqrt{R \delta_R} $ where the amplitude of the pinch is $\delta_R$. 

To go beyond scaling arguments,  we square the mean curvature in Eq.~(\ref{Happrox}) and expand the free energy density as
\begin{align}
f &= \frac{\kappa}{2} \left[ \zeta_{xx}^2+ \frac{\zeta^2}{R_0^4} - \frac{2 \zeta_{xx}}{R_0}+\frac{2 \zeta_{xx} \zeta}{R_0^2} - \frac{2 \zeta}{R_0^3}\right]+ \frac{3}{4}  Y \frac{\zeta^2 }{R_0^2} \label{fexpanded}
\end{align}
The term proportional to $Y$ is the stretching energy due to the strain $u_{yy}$ from the radius differing from the preferred radius $R_0$, with the assumption that $u_{xx}=-u_{yy}$, as was found to be the case for perfect cylinders in Section \ref{sec3}.

The Euler-Lagrange equation associated with Eq.~(\ref{fexpanded}) reads
\begin{align}
0 &= \frac{\delta f}{\delta \zeta} = \kappa \left[ \zeta_{xxxx} + \zeta R_0^{-4} + 2 \zeta_{xx} R_0^{-2} - R_0^{-3} \right] + \frac{3}{2} Y\zeta R_0^{-2} 
\end{align}
\begin{align}
\Rightarrow R_0 &=   R_0^4\zeta_{xxxx}+ 2  R_0^{2}\zeta_{xx} + \zeta \left( 1+ \frac{3}{2} \gamma \right) 
\label{EulerLagrange}
\end{align}
Suppose that the neck is centered on a dislocation at $x=0$. For $x>0$, we have a tubular lattice tessellation where the stretching energy prefers a radius $R_0^+$, and the F\"oppl-von K\'arm\'an number is $\gamma^+ = \tilde \kappa (R_0^+/a)^{-2} $. For $x<0$, the different preferred tubular radius    introduces similar parameters $R_0^-$, $\gamma^-$. We now solve  Eq.~(\ref{EulerLagrange}) for $x>0$, subject to boundary conditions $\zeta'(x\rightarrow\infty)=\zeta''(x\rightarrow\infty)=0$ and $[R_0^+ + \zeta(x\rightarrow\infty)] = R^+_\infty$, where $R^+_\infty=R_0^+\left(1+\left[1+\frac{3}{2} \gamma^+\right]^{-1}\right)\approx R_0^+\left(1+\frac{2}{3} \left(\gamma^+\right)^{-1}\right)$ is the radius of the pristine tubular crystal with this tessellation. Keeping only real solutions, we find
\begin{align}
R(x>0)& = R_0^+ + \zeta(x>0) = R_\infty^+ + c^+ \mathrm{Re}\left[e^{ -x/w+}\right] \label{zetaplus}, \\ 
w^+ &= R_0^+ \left[-1+ i \sqrt{\tfrac{3}{2} \gamma^+}\right]^{-1/2}
\end{align}
where the complex number $w^+$ describes a combination of exponential and oscillatory behavior of $\zeta(x)$, and $c^+$ has yet to be determined.  In the large $\gamma^+$ limit, $R(x>0) \rightarrow R_\infty^+ + c^+ \exp\left(-x/v^+\right)\cos(x/v^+)$ where $v^+ = R_0^+ \left(\frac{3}{2} \gamma^+\right)^{-1/4}$.

\dbedit{At $x=0$, Eq.~\ref{zetaplus} must match onto 
\begin{align}
R(x<0)& = R_0^- + \zeta(x<0) =R_\infty^-+ c^- \mathrm{Re}\left[e^{ x/w^-}\right] , \\
 w^-&= R_0^- \left[-1+ i \sqrt{\tfrac{3}{2}\gamma^-}\right]^{-1/2}  \label{zetaminus}
\end{align}
Ensuring that both $R(x)$ and $R'(x) = \zeta'(x)$ are continuous at $x=0$ requires 
\begin{align}
c^+ &= \left(1+\frac{\mathrm{Re}\left[1/w^+\right]}{\mathrm{Re}\left[1/w^-\right]}\right)^{-1} \left(R_\infty^- - R_\infty^+  \right),  \label{cplus}
\end{align}
and a similar expression for $c^-$ with the $^-$ and $^+$ labels reversed. }

Eqs.~(\ref{zetaplus}-\ref{cplus}) are compared to the numerically calculated radius data in Fig.~\ref{Rvsglide}(a-d) (black curves); our analytic shape profiles provide reasonably good descriptions of the neck provided $\tilde \kappa \gg \tilde \kappa_b$. For the pristine tube radii $R_\infty^-$, $R_\infty^+$, we use radii calculated numerically for defect-free tubes with the associated $(m,n)$; these numerically calculated radii differ slightly ($<1\%$) from the theoretical values $R_\infty = R_0\left[1+\left(1+\frac{3}{2} \gamma\right)^{-1}\right]$ expected from Eq.~(\ref{EulerLagrange}).  (Note that Eq.~(\ref{Rdef}), which defines $R_0$, is exact for a triangular packing of discs in a cylindrical surface, but is only approximate for a tubular crystal of spheres \cite{erickson1973tubular,harris1980tubular}.)  The real part of the complex parameter $w^\pm$ successfully captures the width of the neck, and its imaginary part predicts the oscillatory behavior observed in $R(x)$. As predicted by the scaling argument, for large $\gamma$ the neck width scales as $w\sim \gamma^{-1/4}R_0$. However, even in that limit, $w$ has a complex prefactor resulting in both exponential rise/decay and oscillations in $R(x)$, controlled by the same length-scale. 

If the two dislocations are not far apart compared to $\mathrm{Re}[w] \sim R_0 \gamma^{-1/4}$, then the two calculated neck profiles may overlap significantly. The actual $R(x)$, in order to remain smooth,  must compromise between the two neck profiles;  we  expect this effect leads to small changes in the energetics of dislocation unbinding compared to the calculations in Section \ref{sec3}.

\begin{figure}[htbp]
\centering
\includegraphics[width=0.6\linewidth]{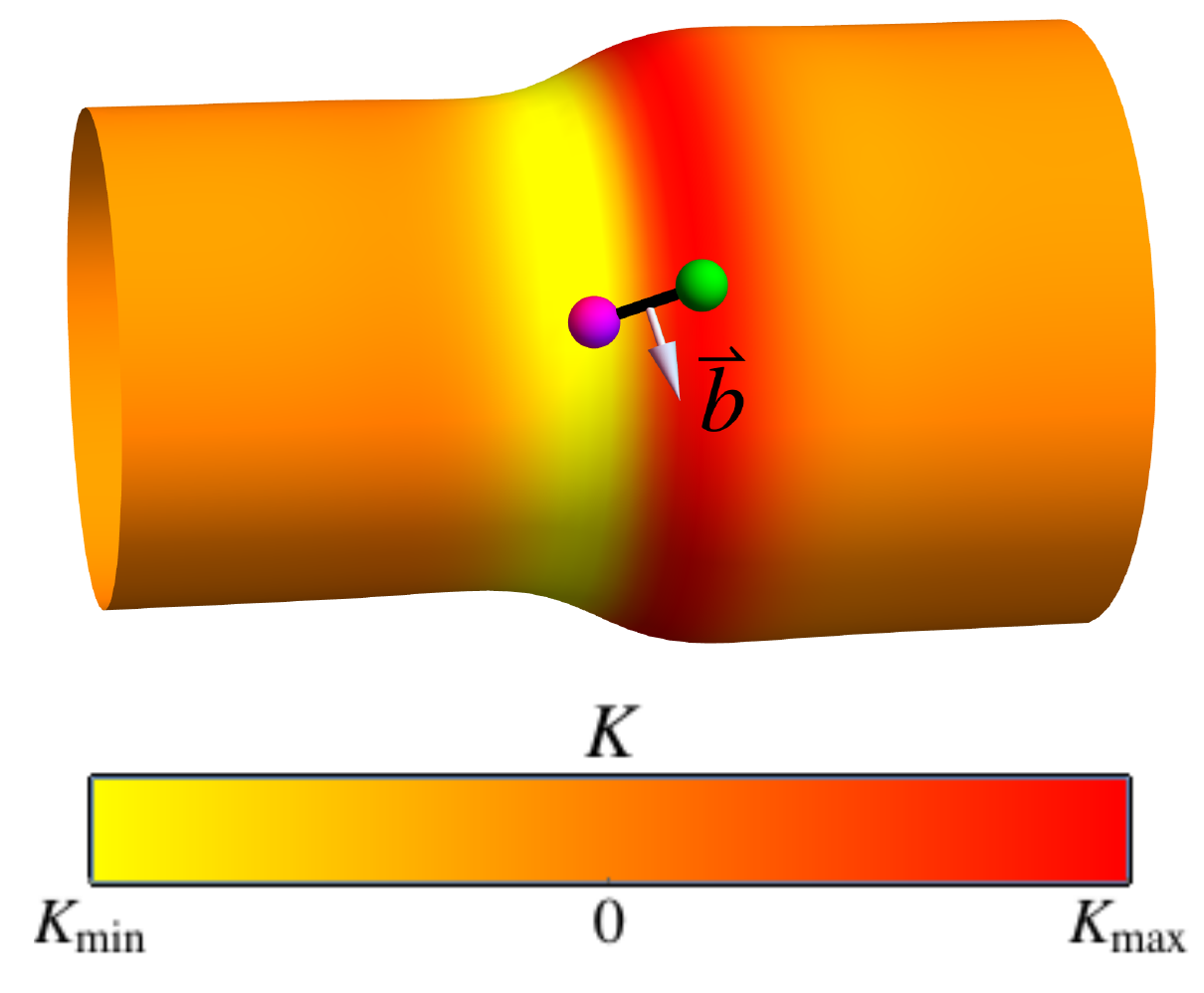}
\caption{Schematic illustration of Gaussian curvature $K$ on a tube with a dislocation. At the neck where the tube radius changes, the Gaussian curvature is mostly positive on the wider side and mostly negative on the narrow side. The dislocation that causes this neck has its fivefold disclination (green sphere) near the ring of maximum $K$ and its sevenfold  disclination (magenta sphere) near the ring of minimum $K$.  \label{Kschemfig}  }
\end{figure}

\dbedit{Examining the Gaussian curvature $K$ near the neck reveals an interesting agreement between the energetically preferred dislocation orientation and the  geometrical view of dislocation motion in tubular crystals.}  Five-coordinated disclinations prefer positive  $K$, while seven-coordinated disclinations prefer negative $K$. Dislocations on Gaussian bumps, for example, prefer to sit near the ring of $K=0$, oriented with the positive disclination closer to the top of the bump where $K>0$, and the negative disclination at larger distance where $K<0$ \cite{vitelucks06}. In our system, a neck where a tube transitions from a larger to a smaller radius has positive $K$ on the wider side of the neck and negative $K$ on the narrower side of the neck, \dbedit{as shown in Fig.~\ref{Kschemfig}}. Thus, a dislocation will to  energetically prefer to orient with the five-coordinated disclination closer to the wider side. This expectation agrees with the geometry of dislocation motion in a tubular crystal: If a dislocation's positive 5-fold disclination is at, say, a larger $x$-coordinate than the negative 7-fold disclination, then the Burgers vector $\mathbf b$ has a negative component along the $\bm{\hat y}$ direction, meaning that the circumference decreases as the dislocation moves in the positive-$x$ direction. \dbedit{This dislocation orientation is shown schematically in Fig.~\ref{Kschemfig}; see also the rightmost dislocation in the simulated tubular crystals of Fig.~\ref{disksfig}(e,f) and in  Fig.~\ref{introfig}(d).} Thus, the positive disclination is closer to the wider side of the tube, and will be on the part of the neck with positive $K$, while the negative disclination is on the narrower side in a region of $K<0$, consistent wih our simulations.

In this paper we have ignored the interactions between dislocations and the Gaussian curvature $K(\mathbf{x})$ that they induce on a tube. These interactions are described by an energy
\begin{align}
F_G &= \frac{Y}{2} \int dA \int dA' \left[S(\mathbf x)-K(\mathbf x)\right] \frac{1}{\Delta^2 _{\mathbf x \mathbf x'}} \left[S(\mathbf x')-K(\mathbf x')\right] \label{FSKcoupling}
\end{align}
where the effective disclination density associated with a dislocation positioned at $\mathbf{x}_b$ is $S(\mathbf x) = \epsilon_{ij} b_i \partial_j \delta(\mathbf{x}, \mathbf{x}_b)$, and $1/\Delta^2_{\mathbf x \mathbf x'}$ is the Green's function of the biharmonic operator \cite{vitelucks06}. The dislocations evidently position themselves near  rings on the tube  where $R(x)$ has an inflection point and the local Gaussian curvature is approximately zero. As the dislocations glide, the necks move with them, such that the contribution of Eq.~(\ref{FSKcoupling}) to the total energy remains  approximately constant. The good agreement between Eqs.~(\ref{zetaplus})-(\ref{cplus}), which neglect the couplings in Eqn.~(\ref{FSKcoupling}), and the numerically calculated neck profile in Fig.~\ref{Rvsglide}(b-d) suggests that the essential role of the dislocations is in switching the preferred radii between $R_0^+$ and $R_0^-$, not their coupling to the Gaussian curvature field.

\section{Conclusions}

We have examined the plastic deformation of tubular crystals by glide of separating dislocation pairs through both continuum elastic calculations and simulations of a discretized harmonic solid sheet with a bending energy wrapped into a tube. Dislocation pairs gliding apart mediate stepwise parastichy transitions, changing the parastichy numbers $(m,n)$ by $\pm 1$ and thus also changing the tube radius $R$ and helical angle $\phi$ according to Eqs.~\ref{Rdef} and \ref{phidef}. The predictions of continuum elasticity calculations offer valuable insights into the numerically calculated mechanics of tubes even when the tube circumference is as small as $\sim 10$ times the lattice spacing. 
		
Tubes under axial elongation stress change their lattice structure to  converge toward the stable $m=n$ achiral state while their radius shrinks. In the regime where the inverse F\"oppl-von K\'arm\'an number $\gamma^{-1}$ is small but finite, the bending modulus $\kappa$ shifts the critical stress required to drive apart dislocations, strengthening  narrow tubes against plastic deformations caused by axial stress. By properly correcting the yield stress $\sigma_{xx}^\dagger$ required to unbind a dislocation pair with a curvature-induced stress offset $-Y \gamma^{-1}=-\kappa/R^2$, we obtain an effective yield stress as a function of $\phi$ that is independent of the $R$. If $\kappa$ is large enough, very small tubes with $R<R_c \propto \sqrt{\kappa/Y}$ may even be unstable to emission of dislocation pairs that widen the tube, driven by the curvature energy alone.  

We have focused on positive extensional stresses because compressional stresses $(\sigma_{xx}<0)$ are complicated by the possibility of an Euler buckling transition. Similar buckling phenomena can complicate the response to a torsional stress $\sigma_{xy}$ of either sign. Both problems are interesting areas for future investigation.

We have also examined how the tube radius changes spatially in the ``neck'' region surrounding a dislocation.  At small reduced bending modulus $\tilde \kappa=\kappa/Ya^2$ the simulated tubular crystals show buckling behavior like that observed in simulated crystalline membranes. For larger $\tilde \kappa$,  the radius varies smoothly as a function of the cylinder's axial coordinate. The neck has width $w\sim \gamma^{-1/4}R_0$ along the tube axis direction and also exhibits damped oscillations in $R(x)$ at the same length-scale, with the dislocation positioned near a ring of zero Gaussian curvature wrapped around the tube. 

Our results suggest several additional avenues for fruitful future investigation. One is to examine parastichy transitions at finite temperature, where dislocation pairs form and unbind at rates depending on an escape energy barrier \cite{amir2013theory}. Another is to consider the response of the tube to bending, which may stabilize dislocations at nonzero, finite separation \cite{amir2014bending}.

Finally,  nontrivial tube conformations could be targeted by the placement of frozen-in defects, as suggested by \dbedit{the numerically observed kinks in the tube axis at the sites of dislocations (Fig.~\ref{kinkfig}).  Patterns of defects are   expected to promote controllable zigzagging and bending tube shapes.} Even more pronounced distortions will be triggered if the disclinations comprising a dislocation separate and migrate to opposite sides of the tube. In the long term,  we hope that improved understanding of the mechanics of dislocation-mediated parastichy transitions will aid in the design of mechanically reconfigurable bulk materials or nanomachines.
\\
\\
\noindent \textbf{Acknowledgements} \\
The authors thank A.\ Amir, E.\ Efrati, and A.\ Azadi, and A.\ Dinsmore for helpful discussions. This work was supported by the National Science Foundation through grant DMR-1306367 and through the Harvard MRSEC Grant DMR-1420570. D.A.B. was supported by Harvard University through the George F. Carrier Postdoctoral Fellowship. This research was also supported in part by the National Science Foundation under Grant No. NSF PHY11-25915. The authors thank the Kavli Institute for Theoretical
Physics for its hospitality while this work was being discussed and prepared for publication.


\begin{thebibliography}{62}
\expandafter\ifx\csname natexlab\endcsname\relax\def\natexlab#1{#1}\fi
\expandafter\ifx\csname bibnamefont\endcsname\relax
  \def\bibnamefont#1{#1}\fi
\expandafter\ifx\csname bibfnamefont\endcsname\relax
  \def\bibfnamefont#1{#1}\fi
\expandafter\ifx\csname citenamefont\endcsname\relax
  \def\citenamefont#1{#1}\fi
\expandafter\ifx\csname url\endcsname\relax
  \def\url#1{\texttt{#1}}\fi
\expandafter\ifx\csname urlprefix\endcsname\relax\def\urlprefix{URL }\fi
\providecommand{\bibinfo}[2]{#2}
\providecommand{\eprint}[2][]{\url{#2}}

\bibitem[{\citenamefont{Adler et~al.}(1997)\citenamefont{Adler, Barabe, and
  Jean}}]{adler1997history}
\bibinfo{author}{\bibfnamefont{I.}~\bibnamefont{Adler}},
  \bibinfo{author}{\bibfnamefont{D.}~\bibnamefont{Barabe}}, \bibnamefont{and}
  \bibinfo{author}{\bibfnamefont{R.~V.} \bibnamefont{Jean}},
  \bibinfo{journal}{\annbot} \textbf{\bibinfo{volume}{80}},
  \bibinfo{pages}{231} (\bibinfo{year}{1997}).

\bibitem[{\citenamefont{Kuhlemeier}(2007)}]{kuhlemeier2007phyllotaxis}
\bibinfo{author}{\bibfnamefont{C.}~\bibnamefont{Kuhlemeier}},
  \bibinfo{journal}{\trendsplantsci} \textbf{\bibinfo{volume}{12}},
  \bibinfo{pages}{143} (\bibinfo{year}{2007}).

\bibitem[{\citenamefont{Pennybacker et~al.}(2015)\citenamefont{Pennybacker,
  Shipman, and Newell}}]{pennybacker2015phyllotaxis}
\bibinfo{author}{\bibfnamefont{M.~F.} \bibnamefont{Pennybacker}},
  \bibinfo{author}{\bibfnamefont{P.~D.} \bibnamefont{Shipman}},
  \bibnamefont{and} \bibinfo{author}{\bibfnamefont{A.~C.}
  \bibnamefont{Newell}}, \bibinfo{journal}{\physicad}  (\bibinfo{year}{2015}).

\bibitem[{\citenamefont{Levitov}(1991{\natexlab{a}})}]{levitov1991phyllotaxis}
\bibinfo{author}{\bibfnamefont{L.~S.} \bibnamefont{Levitov}},
  \bibinfo{journal}{\prl} \textbf{\bibinfo{volume}{66}}, \bibinfo{pages}{224}
  (\bibinfo{year}{1991}{\natexlab{a}}).

\bibitem[{\citenamefont{Levitov}(1991{\natexlab{b}})}]{levitov1991energetic}
\bibinfo{author}{\bibfnamefont{L.~S.} \bibnamefont{Levitov}},
  \bibinfo{journal}{\epl} \textbf{\bibinfo{volume}{14}}, \bibinfo{pages}{533}
  (\bibinfo{year}{1991}{\natexlab{b}}).

\bibitem[{\citenamefont{Douady and Couder}(1992)}]{douady1992phyllotaxis}
\bibinfo{author}{\bibfnamefont{S.}~\bibnamefont{Douady}} \bibnamefont{and}
  \bibinfo{author}{\bibfnamefont{Y.}~\bibnamefont{Couder}},
  \bibinfo{journal}{\prl} \textbf{\bibinfo{volume}{68}}, \bibinfo{pages}{2098}
  (\bibinfo{year}{1992}).

\bibitem[{\citenamefont{Nisoli et~al.}(2009)\citenamefont{Nisoli, Gabor,
  Lammert, Maynard, and Crespi}}]{PhysRevLett.102.186103}
\bibinfo{author}{\bibfnamefont{C.}~\bibnamefont{Nisoli}},
  \bibinfo{author}{\bibfnamefont{N.~M.} \bibnamefont{Gabor}},
  \bibinfo{author}{\bibfnamefont{P.~E.} \bibnamefont{Lammert}},
  \bibinfo{author}{\bibfnamefont{J.~D.} \bibnamefont{Maynard}},
  \bibnamefont{and} \bibinfo{author}{\bibfnamefont{V.~H.}
  \bibnamefont{Crespi}}, \bibinfo{journal}{\prl}
  \textbf{\bibinfo{volume}{102}}, \bibinfo{pages}{186103}
  (\bibinfo{year}{2009}),
  \urlprefix\url{http://link.aps.org/doi/10.1103/PhysRevLett.102.186103}.

\bibitem[{\citenamefont{Nisoli et~al.}(2010)\citenamefont{Nisoli, Gabor,
  Lammert, Maynard, and Crespi}}]{nisoli2010annealing}
\bibinfo{author}{\bibfnamefont{C.}~\bibnamefont{Nisoli}},
  \bibinfo{author}{\bibfnamefont{N.~M.} \bibnamefont{Gabor}},
  \bibinfo{author}{\bibfnamefont{P.~E.} \bibnamefont{Lammert}},
  \bibinfo{author}{\bibfnamefont{J.}~\bibnamefont{Maynard}}, \bibnamefont{and}
  \bibinfo{author}{\bibfnamefont{V.~H.} \bibnamefont{Crespi}},
  \bibinfo{journal}{\pre} \textbf{\bibinfo{volume}{81}},
  \bibinfo{pages}{046107} (\bibinfo{year}{2010}).

\bibitem[{\citenamefont{Erickson}(1973)}]{erickson1973tubular}
\bibinfo{author}{\bibfnamefont{R.~O.} \bibnamefont{Erickson}},
  \bibinfo{journal}{Science} \textbf{\bibinfo{volume}{181}},
  \bibinfo{pages}{705} (\bibinfo{year}{1973}).

\bibitem[{\citenamefont{Lohr et~al.}(2010)\citenamefont{Lohr, Alsayed, Chen,
  Zhang, Kamien, and Yodh}}]{PhysRevE.81.040401}
\bibinfo{author}{\bibfnamefont{M.~A.} \bibnamefont{Lohr}},
  \bibinfo{author}{\bibfnamefont{A.~M.} \bibnamefont{Alsayed}},
  \bibinfo{author}{\bibfnamefont{B.~G.} \bibnamefont{Chen}},
  \bibinfo{author}{\bibfnamefont{Z.}~\bibnamefont{Zhang}},
  \bibinfo{author}{\bibfnamefont{R.~D.} \bibnamefont{Kamien}},
  \bibnamefont{and} \bibinfo{author}{\bibfnamefont{A.~G.} \bibnamefont{Yodh}},
  \bibinfo{journal}{\pre} \textbf{\bibinfo{volume}{81}},
  \bibinfo{pages}{040401} (\bibinfo{year}{2010}),
  \urlprefix\url{http://link.aps.org/doi/10.1103/PhysRevE.81.040401}.

\bibitem[{\citenamefont{Mughal et~al.}(2011)\citenamefont{Mughal, Chan, and
  Weaire}}]{mughal2011phyllotactic}
\bibinfo{author}{\bibfnamefont{A.}~\bibnamefont{Mughal}},
  \bibinfo{author}{\bibfnamefont{H.~K.} \bibnamefont{Chan}}, \bibnamefont{and}
  \bibinfo{author}{\bibfnamefont{D.}~\bibnamefont{Weaire}},
  \bibinfo{journal}{\prl} \textbf{\bibinfo{volume}{106}},
  \bibinfo{pages}{115704} (\bibinfo{year}{2011}).

\bibitem[{\citenamefont{Wood et~al.}(2013)\citenamefont{Wood, Santangelo, and
  Dinsmore}}]{wood2013self}
\bibinfo{author}{\bibfnamefont{D.~A.} \bibnamefont{Wood}},
  \bibinfo{author}{\bibfnamefont{C.~D.} \bibnamefont{Santangelo}},
  \bibnamefont{and} \bibinfo{author}{\bibfnamefont{A.~D.}
  \bibnamefont{Dinsmore}}, \bibinfo{journal}{Soft Matter}
  \textbf{\bibinfo{volume}{9}}, \bibinfo{pages}{10016} (\bibinfo{year}{2013}).

\bibitem[{\citenamefont{Harris}(2009)}]{harris2009carbon}
\bibinfo{author}{\bibfnamefont{P.~J.~F.} \bibnamefont{Harris}},
  \emph{\bibinfo{title}{Carbon nanotube science: synthesis, properties and
  applications}} (\bibinfo{publisher}{Cambridge University Press},
  \bibinfo{year}{2009}).

\bibitem[{\citenamefont{Harris and Erickson}(1980)}]{harris1980tubular}
\bibinfo{author}{\bibfnamefont{W.~F.} \bibnamefont{Harris}} \bibnamefont{and}
  \bibinfo{author}{\bibfnamefont{R.~O.} \bibnamefont{Erickson}},
  \bibinfo{journal}{\jtheorbiol} \textbf{\bibinfo{volume}{83}},
  \bibinfo{pages}{215} (\bibinfo{year}{1980}).

\bibitem[{\citenamefont{Amir et~al.}(2013)\citenamefont{Amir, Paulose, and
  Nelson}}]{amir2013theory}
\bibinfo{author}{\bibfnamefont{A.}~\bibnamefont{Amir}},
  \bibinfo{author}{\bibfnamefont{J.}~\bibnamefont{Paulose}}, \bibnamefont{and}
  \bibinfo{author}{\bibfnamefont{D.~R.} \bibnamefont{Nelson}},
  \bibinfo{journal}{\pre} \textbf{\bibinfo{volume}{87}},
  \bibinfo{pages}{042314} (\bibinfo{year}{2013}).

\bibitem[{\citenamefont{Van~Iterson}(1907)}]{van1907mathematische}
\bibinfo{author}{\bibfnamefont{G.}~\bibnamefont{Van~Iterson}}, Ph.D. thesis,
  \bibinfo{school}{TU Delft, Delft University of Technology}
  (\bibinfo{year}{1907}).

\bibitem[{\citenamefont{Mughal and Weaire}(2014)}]{mughal2014theory}
\bibinfo{author}{\bibfnamefont{A.}~\bibnamefont{Mughal}} \bibnamefont{and}
  \bibinfo{author}{\bibfnamefont{D.}~\bibnamefont{Weaire}},
  \bibinfo{journal}{\pre} \textbf{\bibinfo{volume}{89}},
  \bibinfo{pages}{042307} (\bibinfo{year}{2014}).

\bibitem[{\citenamefont{Mughal et~al.}(2012)\citenamefont{Mughal, Chan, Weaire,
  and Hutzler}}]{mughal2012dense}
\bibinfo{author}{\bibfnamefont{A.}~\bibnamefont{Mughal}},
  \bibinfo{author}{\bibfnamefont{H.~K.} \bibnamefont{Chan}},
  \bibinfo{author}{\bibfnamefont{D.}~\bibnamefont{Weaire}}, \bibnamefont{and}
  \bibinfo{author}{\bibfnamefont{S.}~\bibnamefont{Hutzler}},
  \bibinfo{journal}{\pre} \textbf{\bibinfo{volume}{85}},
  \bibinfo{pages}{051305} (\bibinfo{year}{2012}).

\bibitem[{\citenamefont{Mughal}(2013)}]{mughal2013screw}
\bibinfo{author}{\bibfnamefont{A.}~\bibnamefont{Mughal}},
  \bibinfo{journal}{\philosmag} \textbf{\bibinfo{volume}{93}},
  \bibinfo{pages}{4070} (\bibinfo{year}{2013}).

\bibitem[{\citenamefont{Yakobson}(1998)}]{yakobson1998mechanical}
\bibinfo{author}{\bibfnamefont{B.~I.} \bibnamefont{Yakobson}},
  \bibinfo{journal}{\apl} \textbf{\bibinfo{volume}{72}}, \bibinfo{pages}{918}
  (\bibinfo{year}{1998}).

\bibitem[{\citenamefont{Nardelli
  et~al.}(1998{\natexlab{a}})\citenamefont{Nardelli, Yakobson, and
  Bernholc}}]{PhysRevLett.81.4656}
\bibinfo{author}{\bibfnamefont{M.~B.} \bibnamefont{Nardelli}},
  \bibinfo{author}{\bibfnamefont{B.~I.} \bibnamefont{Yakobson}},
  \bibnamefont{and} \bibinfo{author}{\bibfnamefont{J.}~\bibnamefont{Bernholc}},
  \bibinfo{journal}{\prl} \textbf{\bibinfo{volume}{81}}, \bibinfo{pages}{4656}
  (\bibinfo{year}{1998}{\natexlab{a}}),
  \urlprefix\url{http://link.aps.org/doi/10.1103/PhysRevLett.81.4656}.

\bibitem[{\citenamefont{Nardelli
  et~al.}(1998{\natexlab{b}})\citenamefont{Nardelli, Yakobson, and
  Bernholc}}]{nardelli1998mechanism}
\bibinfo{author}{\bibfnamefont{M.~B.} \bibnamefont{Nardelli}},
  \bibinfo{author}{\bibfnamefont{B.~I.} \bibnamefont{Yakobson}},
  \bibnamefont{and} \bibinfo{author}{\bibfnamefont{J.}~\bibnamefont{Bernholc}},
  \bibinfo{journal}{\prb} \textbf{\bibinfo{volume}{57}}, \bibinfo{pages}{R4277}
  (\bibinfo{year}{1998}{\natexlab{b}}).

\bibitem[{\citenamefont{Yakobson and Avouris}(2001)}]{yakobson2001mechanical}
\bibinfo{author}{\bibfnamefont{B.~I.} \bibnamefont{Yakobson}} \bibnamefont{and}
  \bibinfo{author}{\bibfnamefont{P.}~\bibnamefont{Avouris}}, in
  \emph{\bibinfo{booktitle}{Carbon nanotubes}} (\bibinfo{publisher}{Springer},
  \bibinfo{year}{2001}), pp. \bibinfo{pages}{287--327}.

\bibitem[{\citenamefont{Zhang et~al.}(2009)\citenamefont{Zhang, James, and
  Dumitric{\u{a}}}}]{zhang2009dislocation}
\bibinfo{author}{\bibfnamefont{D.-B.} \bibnamefont{Zhang}},
  \bibinfo{author}{\bibfnamefont{R.}~\bibnamefont{James}}, \bibnamefont{and}
  \bibinfo{author}{\bibfnamefont{T.}~\bibnamefont{Dumitric{\u{a}}}},
  \bibinfo{journal}{\jcm} \textbf{\bibinfo{volume}{130}},
  \bibinfo{pages}{071101} (\bibinfo{year}{2009}).

\bibitem[{\citenamefont{Bettinger et~al.}(2002)\citenamefont{Bettinger,
  Dumitric{\u{a}}, Scuseria, and Yakobson}}]{bettinger2002mechanically}
\bibinfo{author}{\bibfnamefont{H.~F.} \bibnamefont{Bettinger}},
  \bibinfo{author}{\bibfnamefont{T.}~\bibnamefont{Dumitric{\u{a}}}},
  \bibinfo{author}{\bibfnamefont{G.~E.} \bibnamefont{Scuseria}},
  \bibnamefont{and} \bibinfo{author}{\bibfnamefont{B.~I.}
  \bibnamefont{Yakobson}}, \bibinfo{journal}{\prb}
  \textbf{\bibinfo{volume}{65}}, \bibinfo{pages}{041406}
  (\bibinfo{year}{2002}).

\bibitem[{\citenamefont{Huang et~al.}(2006)\citenamefont{Huang, Chen, Ren,
  Wang, Wang, Vaziri, Suo, Chen, and Dresselhaus}}]{huang2006kink}
\bibinfo{author}{\bibfnamefont{J.~Y.} \bibnamefont{Huang}},
  \bibinfo{author}{\bibfnamefont{S.}~\bibnamefont{Chen}},
  \bibinfo{author}{\bibfnamefont{Z.~F.} \bibnamefont{Ren}},
  \bibinfo{author}{\bibfnamefont{Z.~Q.} \bibnamefont{Wang}},
  \bibinfo{author}{\bibfnamefont{D.~Z.} \bibnamefont{Wang}},
  \bibinfo{author}{\bibfnamefont{M.}~\bibnamefont{Vaziri}},
  \bibinfo{author}{\bibfnamefont{Z.}~\bibnamefont{Suo}},
  \bibinfo{author}{\bibfnamefont{G.}~\bibnamefont{Chen}}, \bibnamefont{and}
  \bibinfo{author}{\bibfnamefont{M.~S.} \bibnamefont{Dresselhaus}},
  \bibinfo{journal}{\prl} \textbf{\bibinfo{volume}{97}},
  \bibinfo{pages}{075501} (\bibinfo{year}{2006}).

\bibitem[{\citenamefont{Bozovic et~al.}(2003)\citenamefont{Bozovic, Bockrath,
  Hafner, Lieber, Park, and Tinkham}}]{bozovic2003plastic}
\bibinfo{author}{\bibfnamefont{D.}~\bibnamefont{Bozovic}},
  \bibinfo{author}{\bibfnamefont{M.}~\bibnamefont{Bockrath}},
  \bibinfo{author}{\bibfnamefont{J.~H.} \bibnamefont{Hafner}},
  \bibinfo{author}{\bibfnamefont{C.~M.} \bibnamefont{Lieber}},
  \bibinfo{author}{\bibfnamefont{H.}~\bibnamefont{Park}}, \bibnamefont{and}
  \bibinfo{author}{\bibfnamefont{M.}~\bibnamefont{Tinkham}},
  \bibinfo{journal}{\prb} \textbf{\bibinfo{volume}{67}},
  \bibinfo{pages}{033407} (\bibinfo{year}{2003}).

\bibitem[{\citenamefont{Suenaga et~al.}(2007)\citenamefont{Suenaga,
  Wakabayashi, Koshino, Sato, Urita, and Iijima}}]{suenaga2007imaging}
\bibinfo{author}{\bibfnamefont{K.}~\bibnamefont{Suenaga}},
  \bibinfo{author}{\bibfnamefont{H.}~\bibnamefont{Wakabayashi}},
  \bibinfo{author}{\bibfnamefont{M.}~\bibnamefont{Koshino}},
  \bibinfo{author}{\bibfnamefont{Y.}~\bibnamefont{Sato}},
  \bibinfo{author}{\bibfnamefont{K.}~\bibnamefont{Urita}}, \bibnamefont{and}
  \bibinfo{author}{\bibfnamefont{S.}~\bibnamefont{Iijima}},
  \bibinfo{journal}{\natnano} \textbf{\bibinfo{volume}{2}},
  \bibinfo{pages}{358} (\bibinfo{year}{2007}).

\bibitem[{\citenamefont{Nardelli and Bernholc}(1999)}]{nardelli1999mechanical}
\bibinfo{author}{\bibfnamefont{M.~B.} \bibnamefont{Nardelli}} \bibnamefont{and}
  \bibinfo{author}{\bibfnamefont{J.}~\bibnamefont{Bernholc}},
  \bibinfo{journal}{\prb} \textbf{\bibinfo{volume}{60}},
  \bibinfo{pages}{R16338} (\bibinfo{year}{1999}).

\bibitem[{\citenamefont{Nogales}(2000)}]{nogales2000structural}
\bibinfo{author}{\bibfnamefont{E.}~\bibnamefont{Nogales}},
  \bibinfo{journal}{\annurevbiochem} \textbf{\bibinfo{volume}{69}},
  \bibinfo{pages}{277} (\bibinfo{year}{2000}).

\bibitem[{\citenamefont{Hunyadi et~al.}(2007)\citenamefont{Hunyadi,
  Chr{\'e}tien, Flyvbjerg, and J{\'a}nosi}}]{hunyadi2007microtubule}
\bibinfo{author}{\bibfnamefont{V.}~\bibnamefont{Hunyadi}},
  \bibinfo{author}{\bibfnamefont{D.}~\bibnamefont{Chr{\'e}tien}},
  \bibinfo{author}{\bibfnamefont{H.}~\bibnamefont{Flyvbjerg}},
  \bibnamefont{and} \bibinfo{author}{\bibfnamefont{I.~M.}
  \bibnamefont{J{\'a}nosi}}, \bibinfo{journal}{\biolcell}
  \textbf{\bibinfo{volume}{99}}, \bibinfo{pages}{117} (\bibinfo{year}{2007}).

\bibitem[{\citenamefont{Chr{\'e}tien et~al.}(1992)\citenamefont{Chr{\'e}tien,
  Metoz, Verde, Karsenti, and Wade}}]{chretien1992lattice}
\bibinfo{author}{\bibfnamefont{D.}~\bibnamefont{Chr{\'e}tien}},
  \bibinfo{author}{\bibfnamefont{F.}~\bibnamefont{Metoz}},
  \bibinfo{author}{\bibfnamefont{F.}~\bibnamefont{Verde}},
  \bibinfo{author}{\bibfnamefont{E.}~\bibnamefont{Karsenti}}, \bibnamefont{and}
  \bibinfo{author}{\bibfnamefont{R.~H.} \bibnamefont{Wade}},
  \bibinfo{journal}{\jcb} \textbf{\bibinfo{volume}{117}}, \bibinfo{pages}{1031}
  (\bibinfo{year}{1992}).

\bibitem[{\citenamefont{Nelson}(2012)}]{nelson2012biophysical}
\bibinfo{author}{\bibfnamefont{D.~R.} \bibnamefont{Nelson}},
  \bibinfo{journal}{\annurevbiophys} \textbf{\bibinfo{volume}{41}},
  \bibinfo{pages}{371} (\bibinfo{year}{2012}).

\bibitem[{\citenamefont{Amir and Nelson}(2012)}]{amir2012dislocation}
\bibinfo{author}{\bibfnamefont{A.}~\bibnamefont{Amir}} \bibnamefont{and}
  \bibinfo{author}{\bibfnamefont{D.~R.} \bibnamefont{Nelson}},
  \bibinfo{journal}{\pnas} \textbf{\bibinfo{volume}{109}},
  \bibinfo{pages}{9833} (\bibinfo{year}{2012}).

\bibitem[{\citenamefont{Nelson and Amir}(2013)}]{nelson2013defects}
\bibinfo{author}{\bibfnamefont{D.~R.} \bibnamefont{Nelson}} \bibnamefont{and}
  \bibinfo{author}{\bibfnamefont{A.}~\bibnamefont{Amir}},
  \bibinfo{journal}{arXiv preprint arXiv:1303.5896}  (\bibinfo{year}{2013}).

\bibitem[{\citenamefont{Amir et~al.}(2014)\citenamefont{Amir, Babaeipour,
  McIntosh, Nelson, and Jun}}]{amir2014bending}
\bibinfo{author}{\bibfnamefont{A.}~\bibnamefont{Amir}},
  \bibinfo{author}{\bibfnamefont{F.}~\bibnamefont{Babaeipour}},
  \bibinfo{author}{\bibfnamefont{D.~B.} \bibnamefont{McIntosh}},
  \bibinfo{author}{\bibfnamefont{D.~R.} \bibnamefont{Nelson}},
  \bibnamefont{and} \bibinfo{author}{\bibfnamefont{S.}~\bibnamefont{Jun}},
  \bibinfo{journal}{\pnas} \textbf{\bibinfo{volume}{111}},
  \bibinfo{pages}{5778} (\bibinfo{year}{2014}).

\bibitem[{\citenamefont{Rothen and Koch}(1989)}]{rothen1989phyllotaxis}
\bibinfo{author}{\bibfnamefont{F.}~\bibnamefont{Rothen}} \bibnamefont{and}
  \bibinfo{author}{\bibfnamefont{A.-J.} \bibnamefont{Koch}},
  \bibinfo{journal}{\jphys} \textbf{\bibinfo{volume}{50}}, \bibinfo{pages}{633}
  (\bibinfo{year}{1989}).

\bibitem[{\citenamefont{Nelson}(2002)}]{Nelson}
\bibinfo{author}{\bibfnamefont{D.~R.} \bibnamefont{Nelson}},
  \emph{\bibinfo{title}{Defects and geometry in condensed matter physics}}
  (\bibinfo{publisher}{Cambridge University Press}, \bibinfo{year}{2002}).

\bibitem[{\citenamefont{Landau and Lifshitz}(1975)}]{landau1975elasticity}
\bibinfo{author}{\bibfnamefont{L.~D.} \bibnamefont{Landau}} \bibnamefont{and}
  \bibinfo{author}{\bibfnamefont{E.~M.} \bibnamefont{Lifshitz}},
  \emph{\bibinfo{title}{Elasticity theory}} (\bibinfo{publisher}{Pergamon
  Press}, \bibinfo{year}{1975}).

\bibitem[{\citenamefont{Hirth and Lothe}(1982)}]{hirth1982theory}
\bibinfo{author}{\bibfnamefont{J.~P.} \bibnamefont{Hirth}} \bibnamefont{and}
  \bibinfo{author}{\bibfnamefont{J.}~\bibnamefont{Lothe}},
  \emph{\bibinfo{title}{Theory of dislocations}} (\bibinfo{publisher}{Krieger
  Publishing Company}, \bibinfo{year}{1982}).

\bibitem[{\citenamefont{Seung and Nelson}(1988)}]{seung1988defects}
\bibinfo{author}{\bibfnamefont{H.}~\bibnamefont{Seung}} \bibnamefont{and}
  \bibinfo{author}{\bibfnamefont{D.~R.} \bibnamefont{Nelson}},
  \bibinfo{journal}{\pra} \textbf{\bibinfo{volume}{38}}, \bibinfo{pages}{1005}
  (\bibinfo{year}{1988}).

\bibitem[{\citenamefont{Nelson and Halperin}(1979)}]{nelson1979dislocation}
\bibinfo{author}{\bibfnamefont{D.~R.} \bibnamefont{Nelson}} \bibnamefont{and}
  \bibinfo{author}{\bibfnamefont{B.}~\bibnamefont{Halperin}},
  \bibinfo{journal}{\prb} \textbf{\bibinfo{volume}{19}}, \bibinfo{pages}{2457}
  (\bibinfo{year}{1979}).

\bibitem[{\citenamefont{Bruinsma et~al.}(1982)\citenamefont{Bruinsma, Halperin,
  and Zippelius}}]{bruinsma1982motion}
\bibinfo{author}{\bibfnamefont{R.}~\bibnamefont{Bruinsma}},
  \bibinfo{author}{\bibfnamefont{B.}~\bibnamefont{Halperin}}, \bibnamefont{and}
  \bibinfo{author}{\bibfnamefont{A.}~\bibnamefont{Zippelius}},
  \bibinfo{journal}{\prb} \textbf{\bibinfo{volume}{25}}, \bibinfo{pages}{579}
  (\bibinfo{year}{1982}).

\bibitem[{\citenamefont{Lidmar et~al.}(2003)\citenamefont{Lidmar, Mirny, and
  Nelson}}]{lidmar2003virus}
\bibinfo{author}{\bibfnamefont{J.}~\bibnamefont{Lidmar}},
  \bibinfo{author}{\bibfnamefont{L.}~\bibnamefont{Mirny}}, \bibnamefont{and}
  \bibinfo{author}{\bibfnamefont{D.~R.} \bibnamefont{Nelson}},
  \bibinfo{journal}{\pre} \textbf{\bibinfo{volume}{68}},
  \bibinfo{pages}{051910} (\bibinfo{year}{2003}).

\bibitem[{\citenamefont{Yao and Lordi}(1998)}]{yao1998young}
\bibinfo{author}{\bibfnamefont{N.}~\bibnamefont{Yao}} \bibnamefont{and}
  \bibinfo{author}{\bibfnamefont{V.}~\bibnamefont{Lordi}},
  \bibinfo{journal}{\japplphys} \textbf{\bibinfo{volume}{84}},
  \bibinfo{pages}{1939} (\bibinfo{year}{1998}).

\bibitem[{\citenamefont{Salvetat et~al.}(1999)\citenamefont{Salvetat, Bonard,
  Thomson, Kulik, Forro, Benoit, and Zuppiroli}}]{salvetat1999mechanical}
\bibinfo{author}{\bibfnamefont{J.-P.} \bibnamefont{Salvetat}},
  \bibinfo{author}{\bibfnamefont{J.-M.} \bibnamefont{Bonard}},
  \bibinfo{author}{\bibfnamefont{N.~H.} \bibnamefont{Thomson}},
  \bibinfo{author}{\bibfnamefont{A.~J.} \bibnamefont{Kulik}},
  \bibinfo{author}{\bibfnamefont{L.}~\bibnamefont{Forro}},
  \bibinfo{author}{\bibfnamefont{W.}~\bibnamefont{Benoit}}, \bibnamefont{and}
  \bibinfo{author}{\bibfnamefont{L.}~\bibnamefont{Zuppiroli}},
  \bibinfo{journal}{\apa} \textbf{\bibinfo{volume}{69}}, \bibinfo{pages}{255}
  (\bibinfo{year}{1999}).

\bibitem[{\citenamefont{Lu et~al.}(2009)\citenamefont{Lu, Arroyo, and
  Huang}}]{lu2009elastic}
\bibinfo{author}{\bibfnamefont{Q.}~\bibnamefont{Lu}},
  \bibinfo{author}{\bibfnamefont{M.}~\bibnamefont{Arroyo}}, \bibnamefont{and}
  \bibinfo{author}{\bibfnamefont{R.}~\bibnamefont{Huang}},
  \bibinfo{journal}{\jphysd} \textbf{\bibinfo{volume}{42}},
  \bibinfo{pages}{102002} (\bibinfo{year}{2009}).

\bibitem[{\citenamefont{Wei et~al.}(2012)\citenamefont{Wei, Wang, Wu, Yang, and
  Dunn}}]{wei2012bending}
\bibinfo{author}{\bibfnamefont{Y.}~\bibnamefont{Wei}},
  \bibinfo{author}{\bibfnamefont{B.}~\bibnamefont{Wang}},
  \bibinfo{author}{\bibfnamefont{J.}~\bibnamefont{Wu}},
  \bibinfo{author}{\bibfnamefont{R.}~\bibnamefont{Yang}}, \bibnamefont{and}
  \bibinfo{author}{\bibfnamefont{M.~L.} \bibnamefont{Dunn}},
  \bibinfo{journal}{\nanolett} \textbf{\bibinfo{volume}{13}},
  \bibinfo{pages}{26} (\bibinfo{year}{2012}).

\bibitem[{\citenamefont{{Ko{\v s}mrlj} and
  {Nelson}}(2016)}]{2016PhRvB..93l5431K}
\bibinfo{author}{\bibfnamefont{A.}~\bibnamefont{{Ko{\v s}mrlj}}}
  \bibnamefont{and} \bibinfo{author}{\bibfnamefont{D.~R.}
  \bibnamefont{{Nelson}}}, \bibinfo{journal}{\prb}
  \textbf{\bibinfo{volume}{93}}, \bibinfo{eid}{125431} (\bibinfo{year}{2016}),
  \eprint{1508.01528}.

\bibitem[{\citenamefont{Ding et~al.}(2007)\citenamefont{Ding, Jiao, Wu, and
  Yakobson}}]{ding2007pseudoclimb}
\bibinfo{author}{\bibfnamefont{F.}~\bibnamefont{Ding}},
  \bibinfo{author}{\bibfnamefont{K.}~\bibnamefont{Jiao}},
  \bibinfo{author}{\bibfnamefont{M.}~\bibnamefont{Wu}}, \bibnamefont{and}
  \bibinfo{author}{\bibfnamefont{B.~I.} \bibnamefont{Yakobson}},
  \bibinfo{journal}{\prl} \textbf{\bibinfo{volume}{98}},
  \bibinfo{pages}{075503} (\bibinfo{year}{2007}).

\bibitem[{\citenamefont{Gompper and Kroll}(1996)}]{gompper1996random}
\bibinfo{author}{\bibfnamefont{G.}~\bibnamefont{Gompper}} \bibnamefont{and}
  \bibinfo{author}{\bibfnamefont{D.~M.} \bibnamefont{Kroll}},
  \bibinfo{journal}{\jphysi} \textbf{\bibinfo{volume}{6}},
  \bibinfo{pages}{1305} (\bibinfo{year}{1996}).

\bibitem[{ALG()}]{ALGLIB}
\bibinfo{howpublished}{{http://www.alglib.net}}.

\bibitem[{\citenamefont{Hilbert and Cohn-Vossen}(1952)}]{HilbertGeometry}
\bibinfo{author}{\bibfnamefont{D.}~\bibnamefont{Hilbert}} \bibnamefont{and}
  \bibinfo{author}{\bibfnamefont{S.}~\bibnamefont{Cohn-Vossen}},
  \emph{\bibinfo{title}{Geometry and the Imagination}}
  (\bibinfo{publisher}{Chelsea Publishing Company}, \bibinfo{year}{1952}).

\bibitem[{\citenamefont{Paulose et~al.}(2012)\citenamefont{Paulose,
  Vliegenthart, Gompper, and Nelson}}]{paulose2012fluctuating}
\bibinfo{author}{\bibfnamefont{J.}~\bibnamefont{Paulose}},
  \bibinfo{author}{\bibfnamefont{G.~A.} \bibnamefont{Vliegenthart}},
  \bibinfo{author}{\bibfnamefont{G.}~\bibnamefont{Gompper}}, \bibnamefont{and}
  \bibinfo{author}{\bibfnamefont{D.~R.} \bibnamefont{Nelson}},
  \bibinfo{journal}{\pnas} \textbf{\bibinfo{volume}{109}},
  \bibinfo{pages}{19551} (\bibinfo{year}{2012}).

\bibitem[{\citenamefont{Mahadevan et~al.}(2007)\citenamefont{Mahadevan, Vaziri,
  and Das}}]{mahadevan2007persistence}
\bibinfo{author}{\bibfnamefont{L.}~\bibnamefont{Mahadevan}},
  \bibinfo{author}{\bibfnamefont{A.}~\bibnamefont{Vaziri}}, \bibnamefont{and}
  \bibinfo{author}{\bibfnamefont{M.}~\bibnamefont{Das}},
  \bibinfo{journal}{\epl} \textbf{\bibinfo{volume}{77}}, \bibinfo{pages}{40003}
  (\bibinfo{year}{2007}).

\bibitem[{\citenamefont{Vitelli et~al.}(2006)\citenamefont{Vitelli, Lucks, and
  Nelson}}]{vitelucks06}
\bibinfo{author}{\bibfnamefont{V.}~\bibnamefont{Vitelli}},
  \bibinfo{author}{\bibfnamefont{J.~B.} \bibnamefont{Lucks}}, \bibnamefont{and}
  \bibinfo{author}{\bibfnamefont{D.~R.} \bibnamefont{Nelson}},
  \bibinfo{journal}{\pnas} \textbf{\bibinfo{volume}{103}},
  \bibinfo{pages}{12323} (\bibinfo{year}{2006}).

\bibitem[{\citenamefont{Liu et~al.}(1997)\citenamefont{Liu, Dai, Hafner,
  Colbert, and Smalley}}]{liu1997fullerene}
\bibinfo{author}{\bibfnamefont{J.}~\bibnamefont{Liu}},
  \bibinfo{author}{\bibfnamefont{H.}~\bibnamefont{Dai}},
  \bibinfo{author}{\bibfnamefont{J.~H.} \bibnamefont{Hafner}},
  \bibinfo{author}{\bibfnamefont{D.~T.} \bibnamefont{Colbert}},
  \bibnamefont{and} \bibinfo{author}{\bibfnamefont{R.~E.}
  \bibnamefont{Smalley}}, \bibinfo{journal}{Nature}
  \textbf{\bibinfo{volume}{385}}, \bibinfo{pages}{780} (\bibinfo{year}{1997}).

\bibitem[{\citenamefont{Martel et~al.}(1999)\citenamefont{Martel, Shea, and
  Avouris}}]{martel1999rings}
\bibinfo{author}{\bibfnamefont{R.}~\bibnamefont{Martel}},
  \bibinfo{author}{\bibfnamefont{H.~R.} \bibnamefont{Shea}}, \bibnamefont{and}
  \bibinfo{author}{\bibfnamefont{P.}~\bibnamefont{Avouris}},
  \bibinfo{journal}{Nature} \textbf{\bibinfo{volume}{398}},
  \bibinfo{pages}{299} (\bibinfo{year}{1999}).

\bibitem[{\citenamefont{Katifori}(2008)}]{katifori2008vortices}
\bibinfo{author}{\bibfnamefont{E.}~\bibnamefont{Katifori}}, Ph.D. thesis,
  \bibinfo{school}{Harvard University} (\bibinfo{year}{2008}).

\bibitem[{\citenamefont{Giomi and
  Bowick}(2008{\natexlab{a}})}]{giomi2008elastic}
\bibinfo{author}{\bibfnamefont{L.}~\bibnamefont{Giomi}} \bibnamefont{and}
  \bibinfo{author}{\bibfnamefont{M.~J.} \bibnamefont{Bowick}},
  \bibinfo{journal}{\epje} \textbf{\bibinfo{volume}{27}}, \bibinfo{pages}{275}
  (\bibinfo{year}{2008}{\natexlab{a}}).

\bibitem[{\citenamefont{Giomi and
  Bowick}(2008{\natexlab{b}})}]{giomi2008defective}
\bibinfo{author}{\bibfnamefont{L.}~\bibnamefont{Giomi}} \bibnamefont{and}
  \bibinfo{author}{\bibfnamefont{M.~J.} \bibnamefont{Bowick}},
  \bibinfo{journal}{\pre} \textbf{\bibinfo{volume}{78}},
  \bibinfo{pages}{010601} (\bibinfo{year}{2008}{\natexlab{b}}).

\bibitem[{\citenamefont{Jim{\'e}nez et~al.}(2016)\citenamefont{Jim{\'e}nez,
  Stoop, Lagrange, Dunkel, and Reis}}]{jimenez2016curvature}
\bibinfo{author}{\bibfnamefont{F.~L.} \bibnamefont{Jim{\'e}nez}},
  \bibinfo{author}{\bibfnamefont{N.}~\bibnamefont{Stoop}},
  \bibinfo{author}{\bibfnamefont{R.}~\bibnamefont{Lagrange}},
  \bibinfo{author}{\bibfnamefont{J.}~\bibnamefont{Dunkel}}, \bibnamefont{and}
  \bibinfo{author}{\bibfnamefont{P.~M.} \bibnamefont{Reis}},
  \bibinfo{journal}{\prl} \textbf{\bibinfo{volume}{116}},
  \bibinfo{pages}{104301} (\bibinfo{year}{2016}).

\end{thebibliography}

\appendix{
\dbedit{
\section{Toroidal tubular crystals}
In addition to axial extension and torsion, another type of deformation likely to be encountered by tubular crystals is bending. A uniformly bent rod experiences extension along the outer half and compression along the inner half \cite{landau1975elasticity}. Our focus in this paper on tubes with periodic boundary conditions along the tube axis and under uniformly applied stresses precludes  a full treatment of  dislocation motions in response to general  bending forces. However,  some insights arise from the special case of the torus tessellated with a tubular crystal.  Rather than placing the simulated tubular crystal in a box that is periodic along the $X$ direction, we embed the same bond network topology on a torus in $\mathbb{R}^3$. Any tubular crystal with parastichy numbers $(m,n)$ can be mapped onto a physical torus, provided that the torus's major radius $R_M$ is large enough compared to the tubular crystal's radius $R$ to avoid self-intersection. For example, Fig.~\ref{torusfig}(a) shows a pristine tubular crystal with $(m,n)=(14,18)$ embedded as a torus. Here, $2\pi R_M$ is analogous to the length $L_X$ of the periodic box for cylindrical tubular crystals. Such toroidal crystals are in fact sometimes formed by single-walled carbon nanotubes \cite{liu1997fullerene,martel1999rings}. Rather than imposing external stresses, we observe spontaneous dislocation motion in response to the extensional and compressional strains naturally present in the outer and inner portions of the torus, respectively.}

A strong bending rigidity is necessary to maintain an approximately circular cross-section for the toroidal tubular crystal. Fig.~\ref{torusfig}(b) shows that when the reduced bending rigidity $\tilde \kappa = \kappa /(Ya^2)$ is decreased from 1 to 0.1, the surface distorts into a pinched annulus structure with nearly flat sidewalls. The pinched structure is reminiscent of the instability of pressurized rings to deformations from circular to elliptical conformations \cite{katifori2008vortices}. To study dislocation nucleation and glide on a torus, we hereafter keep the bending rigidity fixed at $\tilde \kappa =1$ to avoid this distortion. 

\dbedit{The extensional strain $u_{xx}$ present in the outer portion of the toroidal tubular crystal creates a stress $\sigma_{xx}$ that should locally favor the appearance of tube-narrowing dislocation pairs. Indeed, we find numerically that even with zero externally imposed stress,  a toroidal tubular crystal can be unstable to the nucleation and glide separation of dislocation pairs at its outer equatorial ring, where $u_{xx}$ is greatest. This maximum strain $u_{xx}$ is controlled by the aspect ratio of the tube, $\mathrm{A.R.}= R_M/R$, as the outer equatorial ring has circumference greater than the circumference $R_M$ of the centerline by a factor $u_{xx} = R/R_M = 1/\mathrm{A.R.}$. In general, as a function of the angular coordinate $v$ shown in Fig.~\ref{torusfig}(a), the axial strain is $u_{xx}(v) = (R/R_M) \cos v$. }

\dbedit{ The simplest application of what we've learned about cylindrical tubular crystals is to predict that dislocation pair nucleation and separation require a strain $u_{xx} \geq \sigma_{c}(\phi)/Y + \tilde \kappa (R_0/a)^{-2}$, where $\sigma_c(\phi)$ is the smallest  critical stress $(\sigma_{xx}-\sigma_{yy})^\dagger_{\text{eff}} $ at a given $\phi$, as shown in Fig.~\ref{sigmadaggertheory}(a) in units of $A=Y/(4\pi)$. Upon setting $u_{xx} = 1/\mathrm{A.R.}$, we obtain for the case of $(m,n)=(14,18)$ a prediction that the aspect ratio must be greater than $\approx 3.26$ in order for dislocations to arise spontaneously. These dislocations will have $\mathbf{b}=\mathbf{a}_1$ for the right-moving dislocation, as is the case in Fig.~\ref{torusfig}(c). Numerically, we find that the critical aspect ratio lies between 2.9 and 3.3. On tori with larger aspect ratios (i.e. larger $R_M$, since fixing $(m,n)$ fixes $R$) the dislocation pairs spontaneously annihilate as soon as they are created by a bond flip.}

 Attempts to create tube-narrowing dislocation pairs along the inside, top, or bottom regions of the torus resulted in immediate defect pair annihilation, as the torus there provides no stress to drive the dislocations apart. Along the inner portion of the torus, we expect that the compressional strain created by the torus geometry would result in formation and separation of tube-\textit{widening} dislocation pairs. However, we did not observe such events in the simulated toroidal tubular crystals due to numerical instabilities.  More realistic models of toroidal crystals would include unbound disclinations in the ground state to screen the local Gaussian curvature \cite{giomi2008elastic,giomi2008defective,jimenez2016curvature}, which would interact in interesting and complex ways with gliding dislocations.

\begin{figure}[htbp]
\centering
\includegraphics[width=\linewidth]{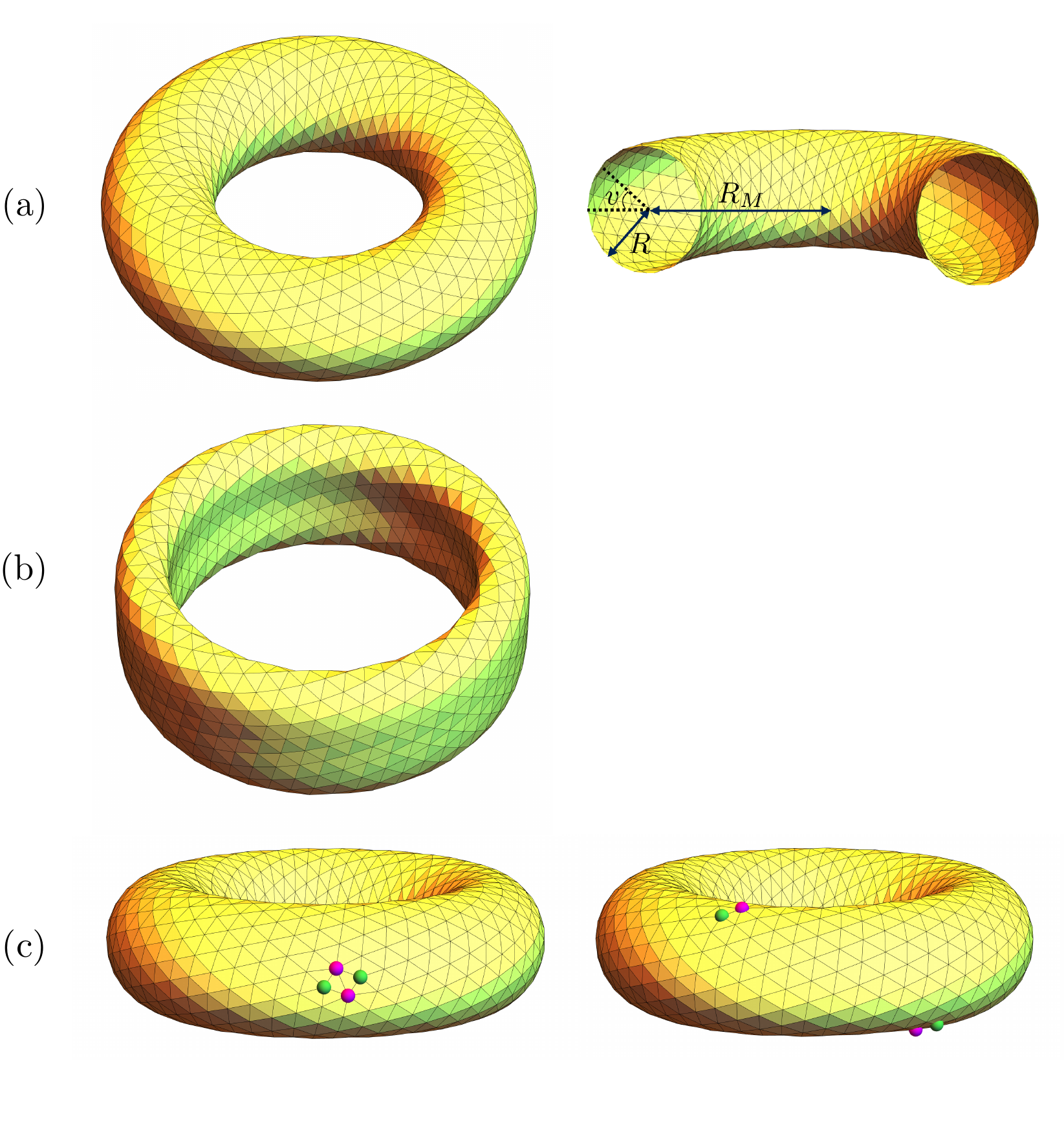}
\caption{ \dbedit{(a,b) Toroidal tubular crystal with $(m,n) = (14,18)$, with reduced bending rigidity $\tilde \kappa = 1$ (a) or $0.1$ (b). Note that pinched sidewalls in (b). Right panel of (a) shows the torus major radius $R_M$, tubular crystal radius $R$, and the angular coordinate $v$. (c) Plastic deformation in the $(m,n)=(14,18)$ tubular crystal with $\tilde \kappa=1$. Left: A dislocation pair whose rightmost member has  $\mathbf{b} = \mathbf{a}_1$ is nucleated near the outer equator. Right: Under zero external stress, the dislocations spontaneously glide apart until coming to rest near the top and bottom of the torus.}    \label{torusfig}}
\end{figure}
}

\end{document}